\documentclass[aps,prapplied,reprint,superscriptaddress,longbibliography,nofootinbib,floatfix]{revtex4-2}

\usepackage{amsmath,amssymb,amsfonts,bm}
\usepackage{graphicx}
\usepackage{booktabs}
\usepackage{hyperref}
\usepackage{xcolor}
\usepackage{xspace}
\colorlet{PURPLE}{purple}
\definecolor{RevisionAdded}{HTML}{0000FF}
\definecolor{RevisionDeleted}{HTML}{FF0000}
\colorlet{REVISIONADDED}{RevisionAdded}
\colorlet{REVISIONDELETED}{RevisionDeleted}
\usepackage[draft]{changes}
  \setaddedmarkup{{\color{RevisionAdded}#1}\xspace}
  \setdeletedmarkup{{\color{RevisionDeleted}#1}\xspace}
\usepackage{todonotes}
\usepackage{algpseudocode}
\usepackage{mathtools}
\usepackage{enumitem}
\graphicspath{{figures/}}

\usepackage[most]{tcolorbox}
\usepackage{algpseudocode}
\newcounter{myalgorithm}

\hypersetup{
  colorlinks=true,
  linkcolor=blue,
  citecolor=blue,
  urlcolor=blue
}
\newcommand{\HHR}[1]{{\color{RevisionAdded}#1}\xspace}

\begin{document}

\title{M\"obius-Guided Diagonal-Gate Compilation with Native Multiqubit Controlled-Phase Gates on Neutral-Atom Processors}

\author{Hairuo Huang}
\affiliation{Beijing Academy of Quantum Information Sciences, Beijing 100193, China}

\author{Yanwu Gu}
\email[]{guyw@baqis.ac.cn}
\affiliation{Beijing Academy of Quantum Information Sciences, Beijing 100193, China}

\author{Chen Huang}
\affiliation{Beijing Academy of Quantum Information Sciences, Beijing 100193, China}
\affiliation{Department of Computer Science and Engineering, The Chinese University of Hong Kong, Hong Kong SAR, China}

\author{Xi Zhao}
\affiliation{Beijing Academy of Quantum Information Sciences, Beijing 100193, China}

\author{Meng-Jun Hu}
\affiliation{Beijing Academy of Quantum Information Sciences, Beijing 100193, China}

\author{Dong E. Liu}
\email[]{dongeliu@mail.tsinghua.edu.cn}
\affiliation{Beijing Academy of Quantum Information Sciences, Beijing 100193, China}
\affiliation{State Key Laboratory of Low Dimensional Quantum Physics, Department of Physics, Tsinghua University, Beijing 100084, China}
\affiliation{Frontier Science Center for Quantum Information, Beijing 100184, China}

\author{Jingbo Wang}
\email[]{wangjb@baqis.ac.cn}
\affiliation{Beijing Academy of Quantum Information Sciences, Beijing 100193, China}

\date{\today}

\begin{abstract}
Diagonal gates are ubiquitous primitives in quantum algorithms, from phase oracles, hypergraph-state preparation, and multi-control logic to Hamiltonian simulation of spin models and digitized lattice field theories, where Ising interactions and local potential terms are diagonal in the encoded basis. Standard compilers, however, often lower diagonal structure into one- and two-qubit gates before neutral-atom hardware can exploit native Rydberg-mediated multiqubit controlled-phase operations. We propose a M\"obius-guided compiler that maps a diagonal phase function to a phase hypergraph via subset-lattice M\"obius inversion. The hypergraph retains the support and angle of each many-body phase term, allowing sparse or local high-order structure to be routed as native multiqubit controlled-phase candidates when feasible and decomposed otherwise. The neutral-atom scheduler accounts for atom motion, interaction-zone constraints, blockade feasibility, and error costs, enabling a direct comparison between native high-order execution and decomposed alternatives. Benchmarks against routed ZAP and ZX-calculus baselines show improved estimated success for algorithmic instances with exploitable three- and four-body phase terms, and comparable performance on predominantly two-body instances. These results provide a feasible compilation strategy for more fully exploiting the native capabilities of neutral-atom hardware, using atom reconfigurability and Rydberg-mediated multiqubit phase operations as practical resources for more efficient quantum computation.
\end{abstract}

\maketitle

\section{Introduction}
\label{sec:intro}

Diagonal gates are ubiquitous primitives in quantum algorithms. They appear as phase oracles in Grover-style search~\cite{Grover1996}, cost-Hamiltonian evolutions in QAOA and related alternating-operator ansatze~\cite{Farhi2014,Wang2018,Hadfield2019}, diagonal time-evolution layers in Hamiltonian simulation and diagonal-unitary synthesis~\cite{Welch2014}, and multiqubit controlled phases that generate hypergraph states~\cite{Rossi2013,Qu2013}. They also arise as multiqubit check or controlled-phase subroutines in coherent-feedback, measurement-free QEC, and stabilizer-readout protocols~\cite{Perlin2023,Hastings2021,Gidney2021,Locher2026}, and as Ising interactions or local potential terms in spin systems~\cite{Morgado2021,Scholl2021,Ebadi2021} and digitized lattice field theories~\cite{Byrnes2006,Jordan2012}. A common feature of these settings is that the target diagonal unitary is often not an arbitrary dense phase function, but a structured collection of sparse local or high-order conditional phase terms. Decomposing such terms immediately into one- and two-qubit gates can obscure the very structure that hardware might implement more directly. This motivates a compilation strategy that preserves diagonal phase structure until the hardware-specific stage, where native multiqubit controlled-phase operations can be selectively exploited.

Existing synthesis methods make diagonal structure accessible through
Pauli-\(Z\) strings, phase polynomials, Walsh transforms, phase gadgets,
CNOT/CZ ladders, and related exact-synthesis constructions
~\cite{Shende2006,Welch2014,Amy2018,Nam2018,Cowtan2020Phase,Kliuchnikov2013,Bocharov2015,Maslov2016}.
Hypergraph-state and SZX-calculus descriptions also represent diagonal
operations through multiqubit phase structure
~\cite{Rossi2013,Qu2013,Carette2019,Carette2020}, and ZX-calculus has recently
been used to synthesize multiqubit controlled-phase gates for neutral-atom
hardware~\cite{Staudacher2024}.  These approaches are powerful, but in many
compilation flows the diagonal layer is eventually lowered to one- and
two-qubit gates before hardware scheduling.  For neutral-atom processors this
early lowering can hide whether a many-body conditional phase can be mapped to a supported multiqubit controlled-phase gate before
scheduling and routing.

Neutral-atom arrays provide a natural platform for preserving diagonal phase
structure during hardware-aware compilation.  Long-lived atomic qubits provide
the information carriers, while Rydberg-mediated interactions supply the
entangling mechanism
~\cite{Jaksch2000,Lukin2001,Saffman2010,Saffman2016,Henriet2020,Browaeys2020,Morgado2021};
optical tweezers enable atom-by-atom assembly and reconfigurable arrays
~\cite{Endres2016,Barredo2016,Barredo2018,Kim2016,Bluvstein2022}; and large
programmable arrays have already reached hundreds of atoms
~\cite{Scholl2021,Ebadi2021}.
Recent experiments have demonstrated controlled-phase gates, parallel
entangling operations, and logical processing with reconfigurable atom arrays
~\cite{Graham2019,Madjarov2020,Evered2023,Bluvstein2024}.
At the same time, emerging architectures increasingly use atom motion,
interaction zones, and readout regions as compilation resources
~\cite{Pichler2018,Baker2021,Cong2022,Nguyen2023,Bluvstein2024,Zhu2025CompilerSurvey}.
These features are particularly relevant for diagonal circuits, in which
each multiqubit support identifies the atoms that must be brought into an
interaction region. Whether retaining that support is advantageous depends on
atom motion, interaction-zone capacity, blockade feasibility, scheduling
conflicts, and the relative cost of a native gate and its decomposition.

The logical multiqubit controlled-phase primitive considered here is the
occupation-projector phase
\begin{equation}
P_S(\theta)
=
\exp\!\left(i\theta\prod_{j\in S} n_j\right),
\qquad
n_j=|1\rangle_j\langle1|.
\label{eq:projector_phase_gate}
\end{equation}
It applies a phase only when all qubits in \(S\) occupy the coupled logical
state.  Thus, the support \(S\) has a dual role: algebraically, it labels a
many-body conditional phase; physically, it identifies the atoms that must
satisfy the geometric and scheduling constraints of a native Rydberg-mediated
operation.

The many-qubit phase structure of a diagonal unitary can be exposed using
M\"obius inversion on the subset lattice~\cite{Rota1964,Aigner1979,Carette2020}.
We use that transform as a neutral-atom compilation principle: a diagonal
phase layer is rewritten as a phase-hypergraph stream whose edges record both
the support and angle of the corresponding occupation-projector phases. Keeping
those hyperedges visible allows a native gate table to determine which
low-degree terms remain multiqubit controlled-phase gates before scheduling and
routing. We compare the resulting M\"obius-native circuit with
ZAP-decomposed~\cite{Huang2026ZAP} and ZX no-insert~\cite{Staudacher2024}
circuits under the same storage-partitioned scheduler, router, and fidelity
model. The routed comparisons cover SAT, QAOA, Ising, hypergraph-state, QRAM,
multiplier, QFT, and GHZ families.

\begin{figure}[!t]
\centering
\includegraphics[width=0.98\columnwidth,height=0.52\textheight,keepaspectratio]{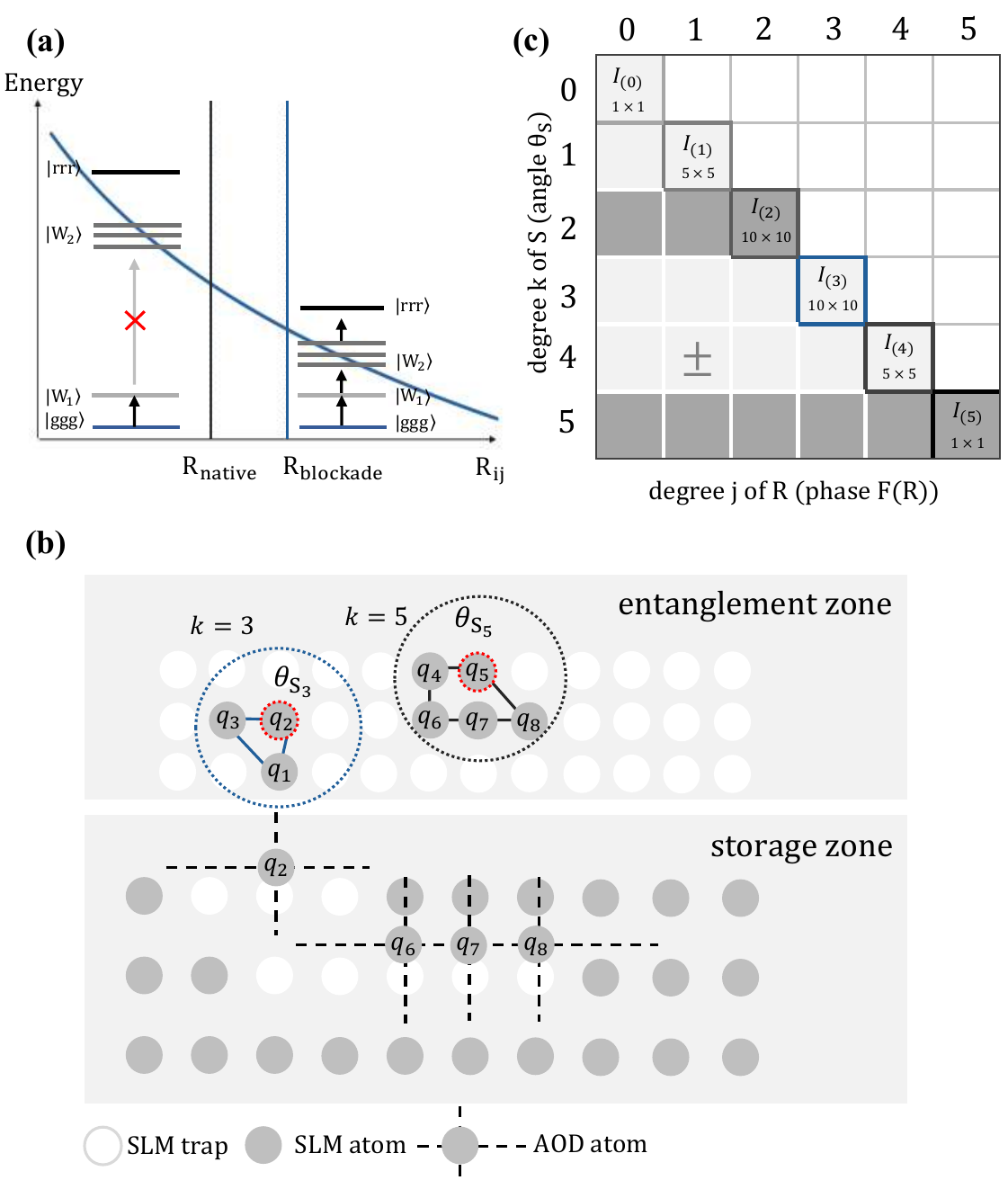}
\caption{\textbf{(a)} Schematic of interaction energies. Variation of the van der Waals frequency shift $V(R_{ij})=C_6/R_{ij}^{6}$ (blue) with interatomic distance $R_{ij}$ for the ground state $|ggg\rangle$, single-excited state $|W_1\rangle$, doubly excited state $|W_2\rangle$, and excited state $|rrr\rangle$. \textbf{(b)} Neutral atom layout with separated storage and computation regions. The figure shows a three-atom support set $S_3=\{q_1,q_2,q_3\}$ ($\theta_{S_3}$, blue) and a five-atom support set $S_5=\{q_4,\dots,q_8\}$ ($\theta_{S_5}$, black). \textbf{(c)} In the degree-graded basis, the M\"obius matrix $M=\zeta^{-1}=\mu_1^{\otimes n}$ takes a block lower-triangular form with identity diagonal blocks $I_{(k)}$ of size $\binom{n}{k}$.}
\label{fig:native_mobius_overview}
\end{figure}

The remainder of the paper is organized as follows. Section~\ref{sec:mobius} introduces the projector-phase primitive, relates it to tunable Rydberg multiqubit controlled-phase gates, and derives the M\"obius phase-hypergraph representation used by the compiler. Section~\ref{sec:benchmark_design} defines the benchmark families and baseline compiler streams used to separate genuinely available many-body diagonal structure from structure lost under early decomposition. Section~\ref{sec:architecture} presents the storage-partitioned neutral-atom routing model and the associated no-fault fidelity estimate. Section~\ref{sec:numerics} reports the numerical comparison across movement cost, execution time, and native-gate error assumptions, including break-even diagnostics for three- and four-qubit controlled phases. Sections~\ref{sec:discussion} and~\ref{sec:conclusion} discuss the resulting hardware implications.

\section{M\"obius phase hypergraphs and their hardware meaning}
\label{sec:mobius}

\subsection{Neutral-atom multiqubit phase gates as projector phases}

Neutral-atom qubits are typically encoded in two long-lived internal states
$|0\rangle$ and $|1\rangle$, while entangling gates use a highly excited
Rydberg state $|r\rangle$ that is optically coupled to one of the logical
states~\cite{Lukin2001,Saffman2010,Saffman2016,Browaeys2020}.  For atoms in a
candidate support $S$, a minimal driven Rydberg model has the form
\begin{align}
H_S(t)=&
\hbar\sum_{j\in S}
\left[\frac{\Omega_j(t)}{2}|r_j\rangle\langle 1_j|+
\frac{\Omega_j^*(t)}{2}|1_j\rangle\langle r_j|\right]
\nonumber\\
&-\hbar\sum_{j\in S}\Delta_j(t)|r_j\rangle\langle r_j|
+\sum_{i<j\in S}V_{ij}|r_ir_j\rangle\langle r_ir_j|,
\label{eq:section_rydberg_hamiltonian}
\end{align}
here \(\Omega_j(t)\) is the Rabi frequency of the laser that drives the transition \(|1_j\rangle\leftrightarrow |r_j\rangle\), and \(\Delta_j(t)\) is the detuning of this drive. The logical state \(|0_j\rangle\) is not coupled by the laser and therefore acts as a dark state. The last term is the Rydberg interaction: if atoms \(i\) and \(j\) are both excited to \(|r\rangle\), the state \(|r_i r_j\rangle\) receives an energy shift \(V_{ij}=C_6/d_{ij}^6\), where \(d_{ij}\) is the distance between the two atoms. As shown in Fig.~\ref{fig:native_mobius_overview}(a), when all atoms in
$S$ are brought within a blockade-compatible geometry, $V_{ij}\gg \Omega_j$
suppresses simultaneous Rydberg excitation.  The laser pulse then drives a
blockaded collective manifold whose dynamical or geometric phase depends on
which atoms are in the optically coupled logical state.  By using resonant~\cite{Yu2022,Stein2025},
off-resonant~\cite{sun2023suppression,Sun2024BufferAtom}, adiabatic~\cite{Pelegri2022}, or echo-compensated pulse sequences~\cite{Levine2019}, the accumulated
diagonal phase can be engineered so that the only uncompensated many-body
phase is applied to the all-occupied computational basis state of the selected
atoms. 

At the compiler level\HHR{,} this calibrated operation is exactly the
occupation-projector phase gate introduced in
Eq.~\eqref{eq:projector_phase_gate}: it adds phase $\theta$ only when every qubit in $S$ is in $|1\rangle$. Thus $P_{\{j\}}(\pi)$ is a $Z$ gate, $P_{\{i,j\}}(\pi)$ is a $CZ$ gate, and $P_{\{i,j,k\}}(\pi)$ is a $CCZ$ gate. More generally, $P_S(\pi)$ is the multi-controlled $Z$ gate on the support $S$, and arbitrary $\theta$ gives a multi-controlled phase rotation. This notation matches the physical trigger condition of a Rydberg multiqubit phase gate. The gate support $S$ identifies the atoms that must be assembled in the entanglement zone; the fan-in $|S|$ determines the required native block size; and the angle $\theta$ is the controlled phase to be synthesized. 

The tunability of $\theta$ has a direct Rydberg-control
interpretation.  In a blockaded manifold, $|0\rangle$ remains dark while a
global laser couples $|1\rangle$ to $|r\rangle$; an $m$-occupied support then contains a symmetric bright-state transition with enhanced coupling
$\sqrt{m}\Omega$, while multiple-Rydberg states are shifted out of resonance. A shaped global detuning pulse $\Delta(\phi,t)$, at fixed Rabi drive, can close the population trajectory and assign a desired relative phase $\phi$ to the all-occupied computational state.  Mohan \emph{et al.} demonstrate smooth, globally addressed pulse families for $C_1P(\phi)$ and $C_2P(\phi)$ gates, namely the two- and three-qubit instances of Eq.~\eqref{eq:projector_phase_gate}, without single-site addressability~\cite{mohan2025parametrized}.  The same blockade-manifold principle extends in principle to larger supports, although pulse optimization, finite-blockade sensitivity, and Rydberg exposure become harder as the fan-in grows. The benchmark below therefore treats $\theta$ as a tunable native parameter only within a conservative low-degree native table and decomposes larger supports.

Fig.~\ref{fig:native_mobius_overview}(b) illustrates this hardware
interpretation for representative three- and five-body supports: each support
is both an algebraic hyperedge and the set of atoms that must be gathered within
the native interaction geometry. In the idealized gate model, correctable
one-body and lower-body phases are calibrated away, absorbed into neighboring
diagonal terms, or represented as additional lower-degree projector phases.
The scheduler receives the logical primitive in
Eq.~\eqref{eq:projector_phase_gate}, while the remaining gate imperfections
enter through the assigned fidelity factor.

\subsection{M\"obius inversion of projector-phase accumulation}
\label{subsec:mobius_projector_accumulation}

At the logical compilation level, a calibrated Rydberg-mediated multiqubit
controlled-phase block is represented by an occupation-projector phase~\eqref{eq:projector_phase_gate}, which applies a phase only when every qubit in the support \(S\) occupies the
coupled logical state \(|1\rangle\).  Let
\begin{equation}
    U_F|T\rangle
    =
    e^{iF(T)}|T\rangle,
    \qquad T\subseteq[n],
\label{eq:diagonal_unitary}
\end{equation}
be a diagonal unitary, where \(|T\rangle\) denotes the computational-basis
state whose occupied set is \(T\).  Equivalently, for a bit string
\(x\in\{0,1\}^n\), \(T(x)=\{j:x_j=1\}\).  We take \(F(T)\) to be a chosen real
representative of the phase; the resulting angles are wrapped modulo \(2\pi\)
when forming the compiled gate list.

The projector phase \(P_S(\theta)\) triggers exactly when \(S\subseteq T\):
\begin{equation}
    P_S(\theta)|T\rangle
    =
    \begin{cases}
      e^{i\theta}|T\rangle, & S\subseteq T,\\
      |T\rangle, & S\nsubseteq T .
    \end{cases}
\label{eq:projector_phase_action}
\end{equation}
Therefore a product of projector phases accumulates phase according to
\begin{equation}
    \prod_{S\subseteq[n]}P_S(\theta_S)|T\rangle
    =
    \exp\!\left(i\sum_{S\subseteq T}\theta_S\right)|T\rangle .
\label{eq:projector_phase_accumulation}
\end{equation}
The \(S=\varnothing\) term is only a global phase and may be omitted whenever
global phases are ignored.  Matching Eq.~\eqref{eq:diagonal_unitary} gives the
subset-zeta relation
\begin{equation}
    F(T)
    =
    \sum_{S\subseteq T}\theta_S,
    \qquad T\subseteq[n].
\label{eq:subset_zeta_relation}
\end{equation}
Its inverse is M\"obius inversion on the subset lattice:
\begin{equation}
    \theta_S
    =
    \sum_{R\subseteq S}
    (-1)^{|S|-|R|}F(R).
\label{eq:thetas}
\end{equation}
Thus \(\theta_S\) is the irreducible conditional phase supported on \(S\):
it is what remains after subtracting all proper-subset contributions from the
basis-state phases.  In the pure projector-phase basis, these coefficients are
fixed uniquely by the diagonal unitary, up to the choice of \(2\pi\) phase
representatives.  A short proof of this subset-lattice inversion, together with
a two-qubit example, is given in Appendix~\ref{app:mobius-inversion}.

Figure~\ref{fig:native_mobius_overview}(c) expresses the relation between computational-basis phases and native projector phases in matrix form. We index the vector $F$ by computational-basis supports $T\subseteq[n]$, with each component $F(T)$ denoting the phase acquired by the computational-basis state whose occupied support is $T$, and index the vector $\theta$ by projector supports $S\subseteq[n]$, with each component $\theta_S$ denoting the phase contributed by the native projector supported on $S$. Because a basis state with support $T$ accumulates the phases of all native projectors with supports $S\subseteq T$, the two vectors satisfy $F=\zeta\theta$, where $\zeta_{T,S}=\mathbf{1}[S\subseteq T]$ is the subset-zeta matrix. It follows that $\theta=MF$, where $M=\zeta^{-1}$ is the corresponding Möbius inversion matrix, with entries $M_{S,R}=(-1)^{|S|-|R|}\mathbf{1}[R\subseteq S]$. When the subsets are ordered by cardinality, $M$ takes a block lower-triangular form: for each support $S$ of degree $k$, the coefficient $\theta_S$ is obtained by inclusion--exclusion over $F(R)$ for all $R\subseteq S$, thereby removing the phase contributions associated with all proper subsets of $S$, while the degree-$k$ diagonal block is an identity matrix $I_{(k)}$ of dimension $\binom{n}{k}$. Appendix~\ref{app:mobius-inversion} provides the subset-lattice and Kronecker-product constructions, a proof of the inversion formula, explicit two- and three-qubit examples, and a discussion of the phase ambiguity modulo $2\pi$.

This projector basis should be distinguished from the conventional Pauli-\(Z\)
parity basis.  Since \(n_j=(I-Z_j)/2\),
\begin{equation}
    \prod_{j\in S} n_j
    =
    2^{-|S|}
    \sum_{R\subseteq S}
    (-1)^{|R|}
    \prod_{j\in R} Z_j .
\label{eq:projector_to_pauli_strings}
\end{equation}
The generator of a single occupation-projector phase therefore expands into
\(2^{|S|}\) commuting Pauli-\(Z\) strings.  Conversely, a Pauli string applies
phase according to parity, rather than the all-occupied trigger condition in
Eq.~\eqref{eq:projector_phase_action}. The M\"obius/projector basis assigns each nonzero coefficient one logical
occupation-projector request, namely a support \(S\) and an angle \(\theta_S\).
During circuit construction, the native gate table fixes which requests remain
multiqubit controlled-phase gates and which are decomposed. Scheduling and
routing evaluate the resulting circuits without changing those choices.

For Boolean phase oracles, this construction reduces to the familiar
algebraic-normal-form or Reed--Muller transform~\cite{Campbell2012} over \(\mathbb F_2\), where
nonzero monomials correspond to \(\pi\)-projector phases.  The real-valued
M\"obius coefficients used here extend the same support-based viewpoint from
Boolean \(\pi\)-phase oracles to general diagonal phase functions with
arbitrary angles; a parallel derivation of this subset-lattice expansion for molecular diagonal Hamiltonian compilation is provided in~\cite{courtney2026oraclefreequantum}. Pauli-string, Walsh, phase-polynomial, and graphical representations provide other useful coordinates for diagonal synthesis ~\cite{Welch2014,Amy2018,Cowtan2020Phase,Carette2020}, but the
projector-phase form exposes the exact logical primitive considered by the
neutral-atom compiler.

\subsection{Phase hypergraph as an intermediate representation for neutral-atom scheduling}
\label{subsec:phase_hypergraph_ir}

The M\"obius decomposition turns a diagonal phase function into a list of projector-phase requests.  Starting from the coefficients in
Eq.~\eqref{eq:thetas}, the compiler first wraps every angle to a canonical representative,
\begin{equation}
    \bar\theta_S
    =
    \operatorname{wrap}_{[-\pi,\pi)}(\theta_S).
    \label{eq:wrapped_coeff}
\end{equation}
This is an exact canonicalization rather than an approximation, since diagonal phases that differ by integer multiples of \(2\pi\) define the same unitary. After wrapping, coefficients that vanish modulo \(2\pi\) are dropped.

The remaining coefficients define a weighted phase hypergraph
\begin{equation}
\begin{aligned}
    \mathcal G_F &= (Q,\mathcal E_F), \qquad Q=[n],\\
    \mathcal E_F
    &=
    \{\, (S,\bar\theta_S):
    \varnothing\neq S\subseteq Q,\;
    \bar\theta_S\not\equiv0\pmod{2\pi}\,\}.
\end{aligned}
\label{eq:hypergraph}
\end{equation}
The empty support \(S=\varnothing\) is omitted because it contributes only a
global phase.  If several terms generate the same support, their angles are
added and wrapped before the edge list is passed to the compiler.

This phase hypergraph is an intermediate representation between algebraic
diagonal synthesis and neutral-atom scheduling.  A hyperedge
\((S,\bar\theta_S)\) records the support and angle of one projector-phase gate
\(P_S(\bar\theta_S)\).  The same support \(S\) is also the physical set of
atoms that would have to be arranged for a native multiqubit controlled-phase
operation. During circuit construction, the native gate table fixes which requests
remain multiqubit controlled-phase gates and which are decomposed. Scheduling
and routing then evaluate the resulting circuit without changing those choices.

This representation also makes explicit when the M\"obius expansion is compact.
For a generic diagonal phase function on \(n\) bits, the edge set
\(\mathcal E_F\) can contain \(2^n-1\) nontrivial hyperedges. Thus the
phase-hypergraph representation is polynomial-size exactly when the nonzero
wrapped M\"obius support is sparse,
\begin{equation}
    |\mathcal E_F|=\operatorname{poly}(n).
    \label{eq:size}
\end{equation}
This is a statement about the size of the resulting representation.  To obtain
the hypergraph efficiently, the phase function must also be given in a form
from which the nonzero M\"obius coefficients can be generated without scanning
all \(2^n\) basis states.

A useful sufficient condition is a local-term description. Suppose
\begin{equation}
    F(x)
    =
    \sum_{\alpha=1}^{N_{\rm loc}} f_\alpha(x_{A_\alpha}),
    \qquad
    |A_\alpha|\le k,
    \label{eq:local_phase_decomposition}
\end{equation}
where \(N_{\rm loc}=\operatorname{poly}(n)\), \(A_\alpha\subseteq[n]\), and each local
term depends only on the bits in \(A_\alpha\).  The M\"obius expansion of
\(f_\alpha\) has support only on nonempty subsets of \(A_\alpha\), and hence
contributes at most \(2^{|A_\alpha|}-1\le 2^k-1\) nontrivial projector-phase
terms. Therefore, before cancellations and modulo-\(2\pi\) wrapping,
\begin{equation}
    |\mathcal E_F|
    \le
    \sum_{\alpha=1}^{N_{\rm loc}}(2^{|A_\alpha|}-1)
    \le
    N_{\rm loc}(2^k-1).
\label{eq:local_complexity}
\end{equation}
The representation is therefore efficient for the constant \(k\), and remains
polynomial-sized for \(k=O(\log n)\).  The locality cancellation and resulting
output-sensitive work bound are proved in
Appendix~\ref{app:local_complexity_proof}. This locality condition captures
the regime relevant to many lattice Hamiltonian terms, including local spin
interactions and digitized field-theory potentials of fixed polynomial degree. Appendix~\ref{app:local_complexity_proof} gives this explicit field-register counting example.
  
The phase-hypergraph representation is architecture independent.
M\"obius inversion outputs projector-phase requests
\((S,\bar\theta_S)\), while the hardware backend assigns their physical
costs. We evaluate the storage-partitioned/shared-zone model of
Sec.~\ref{sec:architecture}, whose costs include motion, zone capacity,
blockade feasibility, scheduling, idle exposure, crosstalk, and native-gate
fidelity; other layouts can apply different geometry, capacity, or relay
constraints~\cite{Huang2026BRIDGE}. The angle \(\bar\theta_S\) distinguishes a \(\pi\)-phase block from a
general controlled-phase rotation. The benchmark uses a degree-dependent
native-error proxy for the \(\pi\)-dominated Boolean-oracle and hypergraph-state
families; calibrated angle- and geometry-dependent errors can replace it. The
native gate table fixes which supports multiqubit gates remain before the
common router and fidelity model evaluate each strategy.

For Boolean phase oracles
\begin{equation}
    U_f|x\rangle=(-1)^{f(x)}|x\rangle ,
    \label{eq:boolean}
\end{equation}
the same construction reduces to the algebraic normal form over
\(\mathbb F_2\):
\begin{equation}
    f(x)
    =
    \bigoplus_{S\subseteq[n]}
    a_S\prod_{j\in S}x_j,
    \qquad
    a_S
    =
    \bigoplus_{R\subseteq S}f(R).
    \label{eq:anf}
\end{equation}
A nonzero degree-\(k\) monomial corresponds to a \(k\)-qubit
\(P_S(\pi)\) request, giving \(Z\), \(CZ\), \(CCZ\), and higher-controlled
\(Z\) gates for \(k=1,2,3,\ldots\).  Thus SAT phase oracles, hypergraph
states, \(p\)-spin Ising terms, QRAM addressing oracles, and multi-control
arithmetic blocks become phase hypergraphs in which high-degree supports remain
visible to the neutral-atom scheduler.  The worked 3-SAT example in
Sec.~\ref{subsec:sat_mechanism} illustrates how literal-controlled Boolean
structure is converted into explicit projector hyperedges before any
one-/two-qubit lowering is applied.

\section{Routed Compiler Model and Benchmark Design}
\label{sec:benchmark_design}

\subsection{Benchmark families}

The benchmark set is organized by the M\"obius degree spectrum rather than by
algorithm name alone.  For a quantum task on $n$ logical qubits, define the retained
degree histogram
\begin{equation}
h_k(n)=\left|\{S:\ |S|=k,\ \bar\theta_S\not\equiv0\pmod{2\pi}\}\right|.
\label{eq:degree_histogram}
\end{equation}

Only $h_3$ and $h_4$ can directly exercise the low-degree native table used in
this work; $h_1$ and $h_2$ are ordinary one- and two-qubit structure, and
$h_{k>4}$ is decomposition pressure under the present hardware assumptions.
Consistent with the tunability discussion above, the four-atom cutoff is a
conservative hardware choice rather than a limitation of the M\"obius
representation: higher-degree supports are retained in the IR but decomposed in
the benchmark. The histogram is computed before routing, so it identifies where
a native opportunity exists; the storage-partitioned scheduler then quantifies
the associated movement, waiting, and crosstalk costs.

\begin{figure*}[!t]
\centering
\includegraphics[width=0.99\textwidth]{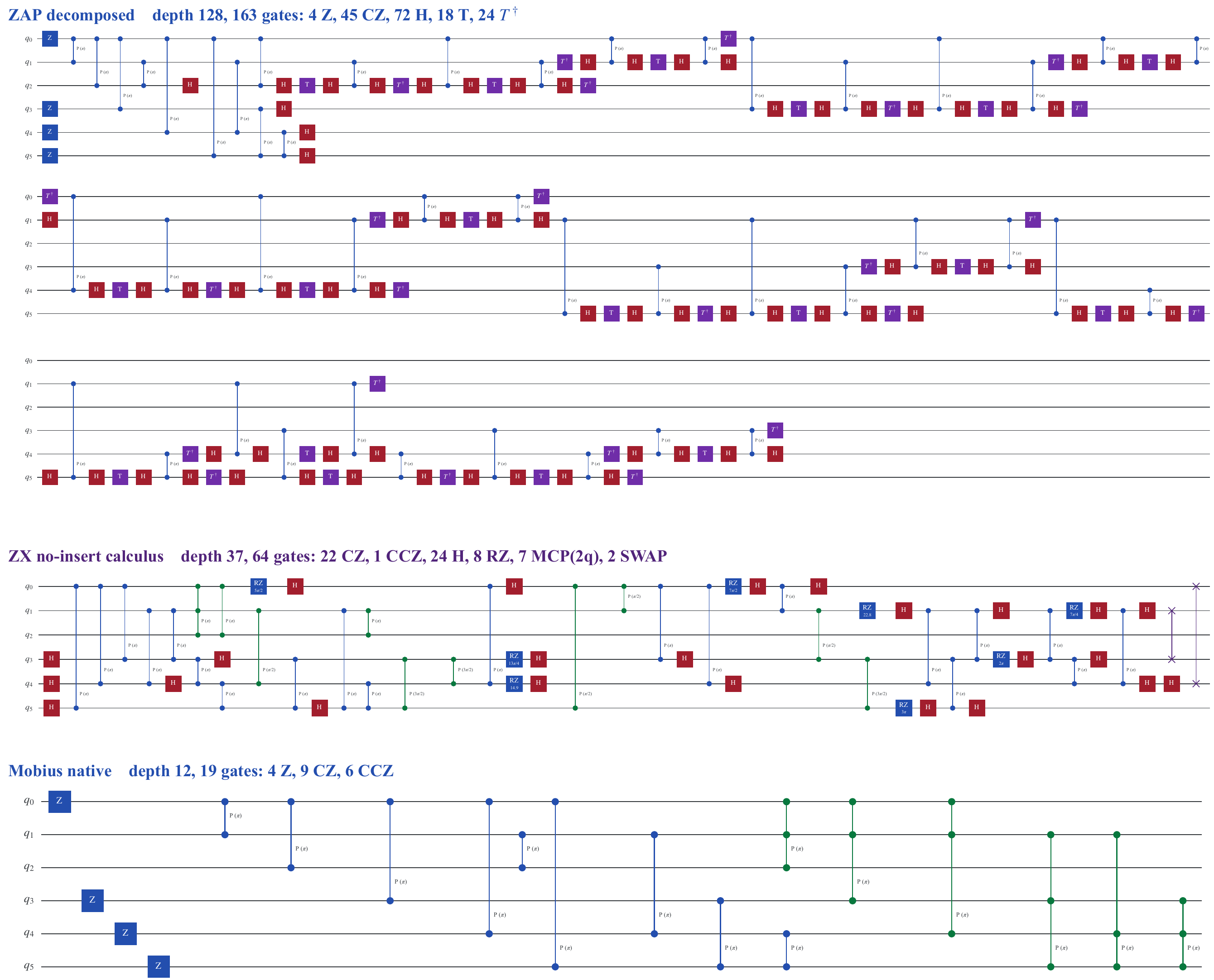}
\caption{Circuit decomposition comparison for the same compact 6-qubit 3-SAT oracle instance. Figs.~\ref{fig:sat3_circuit_comparison}(a)--\ref{fig:sat3_circuit_comparison}(c)
show the ZAP-decomposed, ZX no-insert, and M\"obius-native circuits,
respectively. ZAP lowers the M\"obius projector terms to one- and two-qubit
gates, ZX extracts a different no-insertion phase circuit from the standard
input circuit, and the M\"obius-native path keeps the occupation-projector
supports as native phase gates. In the drawing, single-qubit phases with angle
$\pm\pi/4$ are displayed as $T$ and $T^\dagger$ gates, RZ angles in
Fig.~\ref{fig:sat3_circuit_comparison}(b) are shown modulo $2\pi$ for
compactness, and projector phases are displayed as $Z$ gates or filled-dot
controlled-phase symbols whose dots mark the participating qubit lines and
whose labels give the phase angle $P(\theta)$. Each logical CNOT in the ZAP
template is expanded as $H$--$CZ$--$H$ to match a CZ-native neutral-atom gate
set. The respective displayed depths and gate counts are $128/163$, $37/64$,
and $12/19$. The displayed ZX no-insert circuit contains one $CCZ$ block, while
the M\"obius-native circuit retains six $CCZ$ blocks. The two terminal SWAP
gates in Fig.~\ref{fig:sat3_circuit_comparison}(b) are the output-wire
permutation returned by the ZX extractor; they are retained and counted here so
that all three circuits act on the same fixed qubit ordering.}
\label{fig:sat3_circuit_comparison}
\end{figure*}

The first six benchmark families used in the main scaling plots are chosen to
contain explicit many-body diagonal structure.  The 3-SAT oracle contains
three-literal projector clauses and therefore exposes degree-three native
candidates; it is also the representative circuit used for the decomposition
mechanism in Fig.~\ref{fig:sat3_circuit_comparison}.  The 3-local QAOA benchmark
represents HUBO-style cost Hamiltonians with cubic terms. The $p$-spin Ising
family uses diagonal Hamiltonian terms
$H=\sum_aJ_a\prod_{j\in S_a}Z_j$ with $|S_a|=p$; after conversion from Pauli
parity phases to occupation-projector phases, these instances expose bounded
many-body M\"obius supports. The 4-local hypergraph-state benchmark gives a
clean four-body native-gate stress test. The QRAM oracle and multiplier oracle
represent multi-control addressing and arithmetic blocks, where several
controlled conditions can become high-degree projector phases.

The QFT phase layer and the GHZ-state preparation circuit involve only one- and two-qubit phase or entangling structure. For these algorithms, the Möbius/projector-phase decomposition does not generate genuinely higher-order projector-phase requests; instead, it reduces to the pairwise setting covered by previous compilation approaches. These benchmarks therefore illustrate that, when the input circuit is intrinsically pairwise, the proposed compiler preserves the same effective circuits rather than relying on higher-order native blocks.

\begin{figure}[!t]
\centering
\includegraphics[width=0.98\columnwidth,height=0.36\textheight,keepaspectratio]{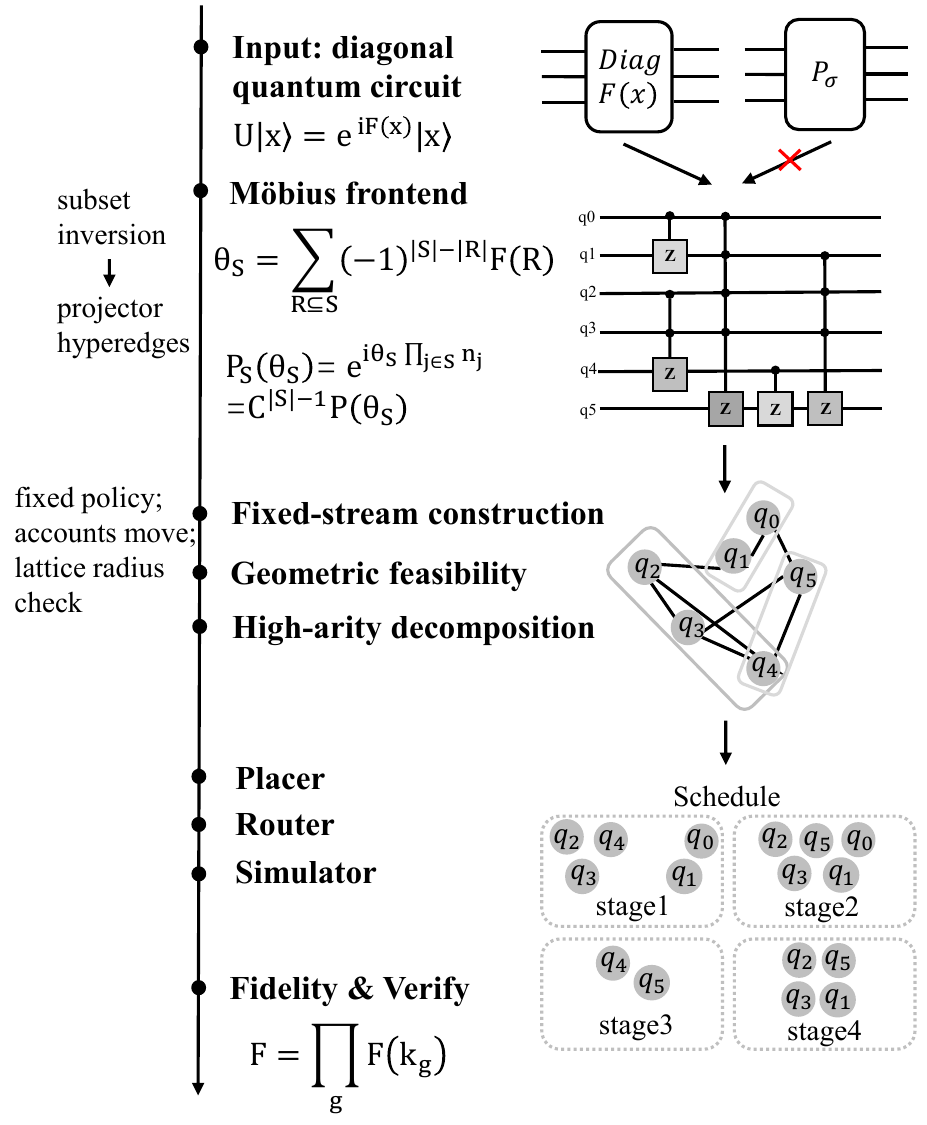}
\caption{The M\"obius Compiler for Atoms pipeline. A diagonal algorithm is Möbius-inverted into projector hyperedges, synthesized as native occupation-projector gates or decomposed (Layer 1), ASAP-scheduled (Layer 2), routed on the storage/entanglement-zoned architecture (Layer 3), and verified bit-exactly (Layer 4). The M\"obius-native and ZAP-decomposed strategies differ only at the synthesis step and share the scheduler, router, and fidelity model.}
\label{fig:moca_pipeline}
\end{figure}

The corresponding compiler organization is summarized in
Fig.~\ref{fig:moca_pipeline}. It separates the algebraic front end, which converts a diagonal phase map into projector hyperedges, from the synthesis,
scheduling, routing, and verification layers used in the numerical benchmark.

\subsection{Baseline compiler streams}
\label{subsec:baselines_fairness}

We compare the M\"obius-native stream with two baselines chosen to separate two effects: architecture-aware neutral-atom routing and preservation of native multiqubit phase structure. The first is a ZAP-decomposed baseline based on the zoned neutral-atom compilation framework of Ref.~\cite{Huang2026ZAP}. ZAP is a strong architecture-level baseline because it models storage zones, entanglement zones, atom movement, and multiplicative fidelity accounting. However, its routed stream is still built from conventional one- and two-qubit operations. It therefore tests whether the proposed advantage comes from keeping high-degree occupation-projector phases visible as native execution opportunities, rather than merely from using a zoned neutral-atom routing model. In this baseline, the same projector terms are lowered before routing: two-body terms remain routed two-qubit phase gates, three-body terms use the exact six-CNOT phase-polynomial template used in the circuit comparison, and larger supports use exact CNOT/phase-gadget expansions. In circuit drawings, CNOTs in the ZAP template are shown as $H$--$CZ$--$H$ to match a CZ-native neutral-atom gate set.

The second baseline is a ZX no-insert baseline following the neutral-atom multi-controlled phase-gate synthesis approach of Ref.~\cite{Staudacher2024}. Unlike ZAP, this baseline does exploit native multiqubit controlled-phase operations, but it exposes them through ZX-calculus synthesis rather than through the M\"obius occupation-projector representation. It therefore tests whether the M\"obius front end improves over another multiqubit-native compilation strategy, not only over one-/two-qubit decomposition. We use the no-insert mode in the main comparison because it gives a conservative ZX baseline and matches the M\"obius-native and ZAP-decomposed unitaries up to global phase in representative noiseless checks.

\subsection{Worked 3-SAT mechanism}
\label{subsec:sat_mechanism}

A compact 3-SAT oracle gives a concrete view of why the M\"obius representation is
useful before routing.  It is a canonical Boolean-oracle for
Grover-style search and QAOA/Max-SAT optimization, and it is the simplest
constraint-satisfaction setting that systematically exposes three-body
conditional structure.  Each clause involves only three variables but may
contain positive and negative literals, so it tests exactly the compiler feature
that this work aims to preserve: whether literal-controlled Boolean logic
remains visible as projector-phase hyperedges rather than being lowered
immediately into one- and two-qubit gates.  Consider a clause
$C=(\ell_a\vee\ell_b\vee\ell_c)$.  The assignment that violates the clause is a
single three-bit pattern: a positive literal is violated by $x_j=0$, while a
negated literal is violated by $x_j=1$.  The corresponding diagonal phase
oracle can therefore be written as a pattern-controlled phase on the three
variables in the clause.  In projector notation the violating-pattern projector
has the form
\begin{equation}
\Pi_C(x)=
\prod_{j\in C_+}(1-x_j)
\prod_{j\in C_-}x_j ,
\label{eq:sat_clause_projector}
\end{equation}
where $C_+$ and $C_-$ denote the variables appearing as positive and negated
literals, respectively.  This expression is already a diagonal local term, but
it still contains negative controls.  Expanding it on the occupation-projector
basis gives
\begin{equation}
\Pi_C(x)=
\sum_{T\subseteq C_+}(-1)^{|T|}
\prod_{j\in C_-\cup T}x_j .
\label{eq:sat_clause_mobius}
\end{equation}
Equivalently, the local M\"obius transform converts the pattern-controlled clause
phase into a short list of ordinary occupation-projector phases.  The largest
support has size three, so the three-literal structure becomes a $CCZ$-type
native candidate rather than a hidden subroutine inside a CNOT/CZ ladder.

This point is not merely a change of notation. Once the three-body support
is visible, the native gate table can retain the clause-level conditional phase
as one multiqubit event. If the support is not admitted, the same hyperedge is
decomposed before scheduling. The common routed evaluator then measures the
movement, idle exposure, and total cost of the resulting strategy. Early
lowering instead replaces the clause by elementary one-/two-qubit structure
before circuit construction can retain it as one three-body projector phase.

Figure~\ref{fig:sat3_circuit_comparison} shows this mechanism for the compact
six-qubit 3-SAT instance used as the representative circuit. Figs.~\ref{fig:sat3_circuit_comparison}(a)--\ref{fig:sat3_circuit_comparison}(c)
are noiselessly equivalent up to global phase, but they expose very different
objects to the neutral-atom routing problem.  ZAP receives a decomposed circuit and
therefore routes many one- and two-qubit gates.  ZX no-insert performs an exact
ZX-calculus synthesis and produces a shorter phase circuit for this instance.
The M\"obius-native path keeps the clause-derived occupation-projector supports,
so six $CCZ$ blocks remain visible as native three-qubit phase candidates.  The
benefit tested later in the paper is precisely whether replacing decomposed
two-qubit ladders by these routed native blocks improves
the total no-fault success after motion,
scheduling, crosstalk, and idle exposure are all included.
The more aggressive ZX insert variant for the same instance is shown in
Appendix~\ref{app:zx_insert_diagnostic}; we keep the no-insert variant in the
main figure because it is the conservative ZX baseline used throughout the
routed benchmark.
The same logic extends from 3-SAT to the many-body benchmark families in this
section: 3-local QAOA and $p$-spin Ising expose few-body Hamiltonian terms,
hypergraph states expose explicit hyperedges, and QRAM or multiplier oracles
expose multi-control addressing or arithmetic conditions.

\section{Storage-partitioned neutral-atom architecture and routed fidelity estimate}
\label{sec:architecture}

\subsection{Concrete architecture model}

We use a zoned neutral-atom platform as the execution model. As shown in
Fig.~\ref{fig:native_mobius_overview}(b),
qubit atoms are initially placed in storage regions,
while entangling gates are performed in a shared entanglement zone. For each
operation generated by the compiler, whether it acts on one, two, or multiple
qubits, the router decides where the operation should be executed. It then
moves any required atoms to that location if they are not already there, adds
the necessary activation and deactivation steps, and schedules both the atom
movements and gate operations in time.

The architectural constraints therefore enter the numerical results through the
routed operation sequence.  The shared zone limits how many atoms can participate
in a native block, the native-block radius $R_{\rm nat}$ tests whether those
atoms can form a feasible local clique, the movement-throughput limit controls
how many transport operations can overlap, and qubit conflicts prevent supports
sharing an atom from executing simultaneously.  After routing, these constraints
have been converted into observable quantities: scheduled stages, movement
events and distances, transfer events, total duration, idle exposure, and
crosstalk exposure. These are the quantities that enter the independent-factor fidelity estimate
below.

All native blocks compete for the shared entanglement zone.  Two scheduled operations conflict if they share atoms, and native blocks additionally conflict through active-zone occupancy and finite movement throughput.  Thus the reported stage count is a scheduled makespan proxy, not merely a gate-count depth.

\subsection{Independent-factor fidelity
estimate}

The main numerical results do not use a closed-form local objective to
choose between native and decomposed realizations. Each strategy first produces a concrete gate-support stream. That stream
is scheduled and routed on the same storage/entanglement-zone architecture,
and the modeled gate, transfer, idle, and coherence factors are multiplied
along the resulting operation sequence:
\begin{equation}
\begin{aligned}
P_0
=&\,F_{1q}^{N_{1q}}F_{2q}^{N_{2q}}
\prod_{g\in\mathcal G_{\rm nat}}F_{\rm nat}^{(|g|)}
F_{\rm transfer}^{N_{\rm transfer}}F_{\rm idle}^{N_{\rm idle}}
\\
&\times
\exp\!\left[-\frac{1}{T_2}
\sum_{q=1}^{n}\left(T_{\rm tot}-T_q^{\rm busy}\right)\right].
\end{aligned}
\label{eq:routed_no_fault}
\end{equation}
Here $N_{1q}$ and $N_{2q}$ count elementary gates after routing,
$\mathcal G_{\rm nat}$ is the set of routed native multiqubit operations,
$N_{\rm transfer}$ counts activation and deactivation events, and
$N_{\rm idle}$ denotes idle atom-layer exposure during entangling operations.
The routed duration is $T_{\rm tot}$, and $T_q^{\rm busy}$ is the accumulated
time during which qubit $q$ participates in a routed operation.  Both quantities
are circuit- and strategy-dependent outputs of the scheduler, not calibrated
hardware constants like $T_2$; hence they are not listed in
Table~\ref{tab:params}.  The storage partitions, shared-zone capacity, movement
throughput, and native-radius test do not appear as separate factors in
Eq.~\eqref{eq:routed_no_fault}; instead, they determine the routed values of
$N_{\rm transfer}$, $N_{\rm idle}$, $T_{\rm tot}$, and $T_q^{\rm busy}$. Atom motion therefore affects the \(P_0\) through the inserted
transfer events and through the duration-dependent coherence factor. The routed
fidelity model does not include a separate distance-dependent motion factor.
Equation~\eqref{eq:routed_no_fault} defines \(P_0\), the independent-factor
estimate of routed circuit fidelity used to compare strategies under common
parameters. Device prediction would require jointly calibrated noise, loss,
and motion models.

\subsection{Native-gate and hardware parameters}

Table~\ref{tab:params} lists the timing, fidelity, and native-routing
parameters used in the routed benchmark. The one-/two-qubit, transfer, idle, coherence, movement, and native-gate entries are chosen at the level of recent reconfigurable neutral-atom processors, where local single-qubit fidelities near $99.91\%$ and two-qubit fidelities near $99.5$--$99.55\%$ have been reported~\cite{Bluvstein2024}. The native three-/four-qubit entries are conservative architecture-level assumptions, not experimental claims: optimized controlled-controlled-phase and multiqubit Toffoli proposals report higher theoretical fidelities than the benchmark values used here~\cite{mohan2025parametrized,Yu2022}. More generally, channel-spectrum benchmarking provides a scalable route for
extracting process fidelity, stochastic fidelity, and unitary parameters of
target unitary gates or circuit fragments from their noisy spectra~\cite{Gu2023CSB}.  Such measurements would be a natural way to replace
the native-gate parameters used here by calibrated data for diagonal
multiqubit modules.

The benchmark native table assigns representative native-gate entries only to
low-degree multiqubit projector-phase modules beyond ordinary controlled-phase
gates.  Native feasibility is then checked through the entanglement-zone
capacity and the pairwise geometry predicate.  Two-qubit controlled phases
remain in the ordinary two-qubit gate table; the compiler-level use of the
native table and the fallback for unsupported supports are described in
Sec.~\ref{sec:numerics}.

\begin{table*}[t]
\caption{Numerical parameters used in the routed timing and no-fault fidelity estimates. The timing, fidelity, decoherence,
motion, and native-feasibility entries define the architecture-level model used
for all strategies. One- and two-qubit fidelities are chosen at the level of
recent neutral-atom processors~\cite{Bluvstein2024};
the native three- and four-qubit fidelities are representative
assumptions~\cite{mohan2025parametrized,Yu2022} and can be replaced by calibrated
device data.}
\label{tab:params}
\begin{ruledtabular}
\begin{tabular}{llll}
Block & Parameter & Value & Role \\
\hline
Timing & $t_{1q}$ & $52~\mu\mathrm{s}$ & one-qubit gate duration \\
 & $t_{2q}$ & $0.36~\mu\mathrm{s}$ & two-qubit gate duration \\
 & $t_{\rm multiq}$ & $0.576~\mu\mathrm{s}$ & native multiqubit gate duration, $1.6t_{2q}$ \\
 & $t_{\rm transfer}$ & $15~\mu\mathrm{s}$ & atom activate/deactivate duration \\
\hline
Fidelity factors & $F_{1q}$ & $0.9997$ & one-qubit gate fidelity \\
 & $F_{2q}$ & $0.995$ & two-qubit gate fidelity \\
 & $F_{\rm transfer}$ & $0.999$ & atom-transfer fidelity \\
 & $F_{\rm idle}$ & $0.9975$ & crosstalk factor, $1-(1-F_{2q})/2$ \\
 & $F_{\rm nat}^{(3)}$ & $0.981557$ & native three-qubit gate fidelity \\
 & $F_{\rm nat}^{(4)}$ & $0.968852$ & native four-qubit gate fidelity \\
\hline
Decoherence and motion & $T_2$ & $1.5\times10^6~\mu\mathrm{s}$ & idle decoherence time \\
 & movement duration & $200\sqrt{d/110}~\mu\mathrm{s}$ & BigMove/Park duration, distance $d$ in $\mu$m \\
\hline
Native feasibility & $k_{\rm nat}^{\max}$ & $4$ & supports with $3\le |S|\le4$ may stay native \\
 & $R_{\rm nat}$ & $8~\mu\mathrm{m}$ & pairwise native-block clique radius \\
\end{tabular}
\end{ruledtabular}
\end{table*}

The sensitivity model scans the assumed three- and four-qubit native error
probabilities, \(p_3=1-F_{\rm nat}^{(3)}\) and
\(p_4=1-F_{\rm nat}^{(4)}\), while holding the routed one-/two-qubit, transfer,
idle, and movement parameters fixed. The resulting phase diagram identifies
where M\"obius-native has larger \(P_0\) than each baseline. The native-gate
entries and the \((p_3,p_4)\) sweep summarize the assumed contribution of
microscopic Rydberg error channels.

\section{Numerical
implementation and results}
\label{sec:numerics}

\subsection{Benchmark
implementation}

The numerical benchmark is organized as a strict shared-model comparison.
For each benchmark family, qubit count $n$, partition count, and random seed, we
first construct a diagonal phase algorithm.  The M\"obius step is applied only to
these diagonal phase blocks, such as phase oracles, cost-Hamiltonian
evolutions, and controlled-phase layers.  Non-diagonal gates that surround such
blocks in full algorithms are not transformed by M\"obius inversion; when
included in an algorithm, they are kept as ordinary circuit operations and routed
by the same neutral-atom scheduler.  Most benchmark families are already
written as occupation-projector supports.

Algorithm~\ref{alg:mobius_native} is the procedural summary of the M\"obius-native path used in the
routed benchmark.  It should be read as the interface between the diagonal front
end and the common neutral-atom backend.  First, the algorithm constructs a
local computational-basis phase table for each diagonal clause generated by the
benchmark family.  Occupation-projector clauses have phase only on the
all-ones entry, while bit-pattern oracle clauses have phase only on the specified pattern.

Second, Algorithm~\ref{alg:mobius_native} applies the subset M\"obius inversion locally on each clause
support.  The subset-zeta relation in Eq.~\eqref{eq:subset_zeta_relation},
restricted to the local set $A$, is inverted by Eq.~\eqref{eq:thetas}.  This
step produces projector requests $(S,\theta_S)$ with $S\subseteq A$; requests
from all clauses are then accumulated, equal supports are merged, phases are
wrapped modulo $2\pi$, and zero-angle terms are discarded.  Thus the M\"obius
support set is fixed by the diagonal phase table, up to the chosen $2\pi$
representatives; the backend cost model does not choose the M\"obius
coefficients.

Third, Algorithm~\ref{alg:mobius_native} exposes this fixed projector-phase list to the gate set used by the routed comparison.  One- and two-qubit terms enter the ordinary elementary
gate stream.  Multiqubit supports within the native table are kept as Rydberg
projector-phase candidates, while unsupported larger supports are not searched
over alternative mixtures of three- and four-qubit blocks in this benchmark.
They are lowered by the same fixed CNOT/phase decomposition template used by the
decomposed baseline, and only the resulting one-/two-qubit elementary supports
are passed to the router.  The final lines of Algorithm~\ref{alg:mobius_native} then run the common
ASAP scheduler, storage/entanglement-zone router, and \(P_0\)
evaluation. In the main plotted benchmark ensemble the retained M\"obius
supports have degree at most four, so the unsupported-support branch is a
conservative fallback rather than the source of the reported native advantage.

Given the support stream defined by Algorithm~\ref{alg:mobius_native}, the three compiler streams
differ only in what they hand to the same routed evaluator.  The
M\"obius-native stream preserves the native candidates selected by the table
exposure step.  The ZAP baseline intentionally removes this native opportunity
by expanding every projector phase into one-/two-qubit gates before routing.
The ZX no-insert baseline starts from the corresponding source circuit, applies
ZX-calculus simplification without insertion, and extracts the resulting support
stream; the insert-mode ZX variant is therefore reported separately as a
stronger front-end diagnostic rather than as the fixed routed baseline.

Figure~\ref{fig:moca_pipeline} gives the module-layer view of the workflow.
Once a support stream has been formed, all strategies use the same greedy
scheduler, storage and entanglement-zone router, and \(P_0\) model from
Sec.~\ref{sec:architecture}; the purpose is a reproducible shared-model
comparison rather than a globally optimal synthesis or routing claim.

\begin{tcolorbox}[
  enhanced,
  breakable,
  width=\columnwidth,
  colback=white,
  frame hidden,
  boxrule=0pt,
  left=0pt,
  right=0pt,
  top=3pt,
  bottom=3pt,
  borderline north={0.8pt}{0pt}{black},
  borderline south={0.8pt}{0pt}{black},
]
\refstepcounter{myalgorithm}
\noindent\textbf{Algorithm 1} M\"obius-to-native phase-hypergraph compilation
\label{alg:mobius_native}
\vspace{2pt}
\hrule
\vspace{3pt}
\noindent\textbf{Input:} Algorithm family $a$; qubit count $n$; partition count $N_P$; seed $s$; architecture $\mathcal A$; native cutoff $k_{\rm nat}^{\max}=4$.

\noindent\textbf{Output:} M\"obius-native support stream, routed operation
sequence, and routed metrics.
\begin{algorithmic}[1]
\State Set partition size $p\gets\max\{3,\lceil n/N_P\rceil\}$.
\State Generate diagonal phase clauses for algorithm family $a$: occupation-projector supports or local bit-pattern phase oracles.
\For{each local diagonal clause on qubits $A$}
    \State Build its computational-basis phase table $F_A(R)$ for $R\subseteq A$~\eqref{eq:subset_zeta_relation}.
    \State Apply subset M\"obius inversion to obtain projector terms $(S,\theta_S)$ with $S\subseteq A$~\eqref{eq:thetas}.
\EndFor
\State Canonicalize phases modulo $2\pi$, merge duplicate supports, and discard zero phases, giving $\mathcal T=\{(S_j,\theta_j)\}$~\eqref{eq:hypergraph}.
\State Record the degree histogram $h_k=|\{j:|S_j|=k\}|$ and other M\"obius statistics.
\State Initialize the M\"obius-native support stream $G_{\rm M}\gets[\,]$.
\For{each term $(S_j,\theta_j)\in\mathcal T$}
    \If{$|S_j|=1$}
        \State Append the corresponding one-qubit phase support to $G_{\rm M}$.
    \ElsIf{$|S_j|=2$}
        \State Append the corresponding two-qubit controlled-phase support to $G_{\rm M}$.
    \ElsIf{$3\le |S_j|\le k_{\rm nat}^{\max}$}
        \State Append $S_j$ as one native multiqubit projector-phase support.
    \Else
        \State Lower $P_{S_j}(\theta_j)$ with the fixed CNOT/phase decomposition template.
        \State Append only the resulting one-/two-qubit elementary supports to $G_{\rm M}$.
    \EndIf
\EndFor
\State Count one-, two-, and multiqubit supports in $G_{\rm M}$.
\State Run the common ASAP scheduler on $G_{\rm M}$.
\State Route every scheduled stage on the same storage/entanglement-zone architecture $\mathcal A$.
\State Construct the routed operation sequence, including initialization, activation, deactivation, motion, parking, gate, and crosstalk events.
\State Evaluate the operation sequence using the common fidelity table and Eq.~\eqref{eq:routed_no_fault}.
\State Return stages, movement events, movement distance, movement duration,
total duration, and \(P_0\).
\end{algorithmic}
\end{tcolorbox}

\begin{figure*}[!t]
\centering
\includegraphics[width=0.88\textwidth]{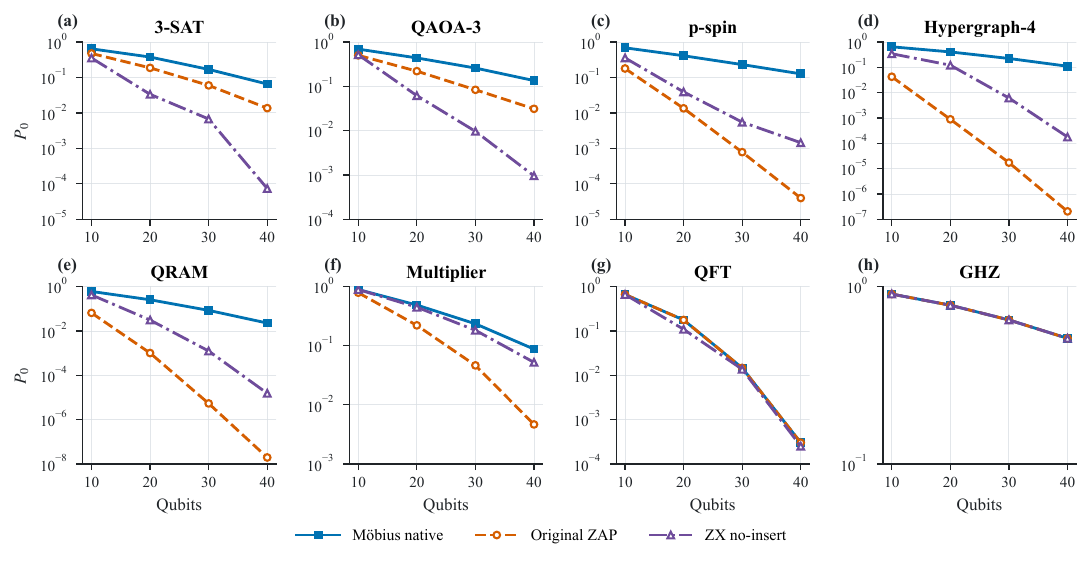}
\caption{\(P_0\) for eight routed benchmark families.
Figs.~\ref{fig:eight_panel_fidelity}(a)--\ref{fig:eight_panel_fidelity}(h)
compare the M\"obius-native, ZAP-decomposed, and ZX no-insert circuits on the
same benchmark instances and logarithmic scale.
Figs.~\ref{fig:eight_panel_fidelity}(a)--\ref{fig:eight_panel_fidelity}(f)
contain explicit many-body Hamiltonian or oracle terms;
Figs.~\ref{fig:eight_panel_fidelity}(g) and~\ref{fig:eight_panel_fidelity}(h)
are the QFT and GHZ two-body controls. Equation~\eqref{eq:routed_no_fault}
defines the plotted quantity.}
\label{fig:eight_panel_fidelity}
\end{figure*}
\begin{figure*}[!t]
\centering
\includegraphics[width=0.88\textwidth]{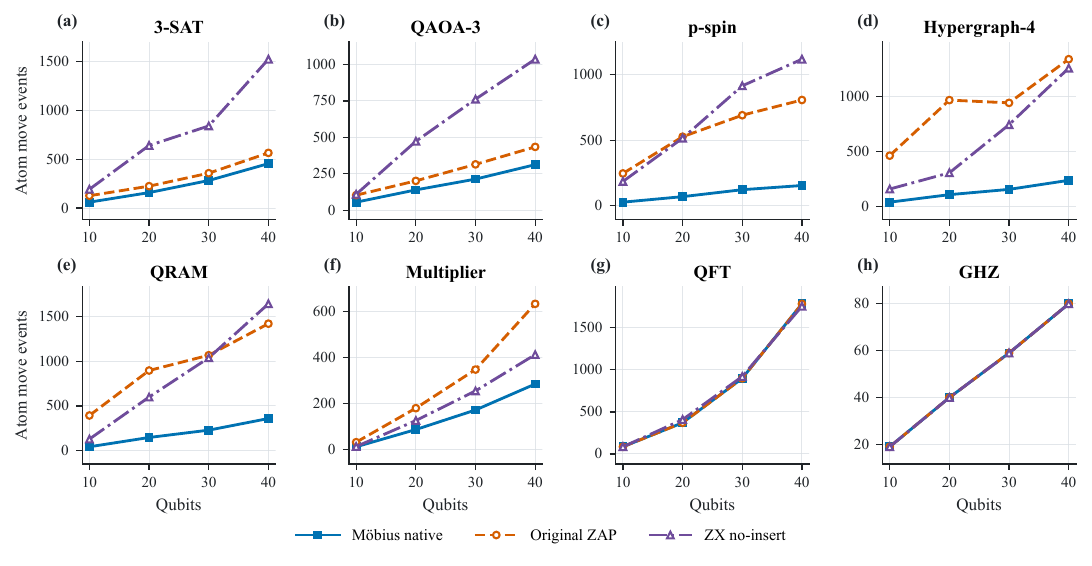}
\caption{Atom-movement comparison for eight common algorithmic benchmark
families. Figs.~\ref{fig:eight_panel_moves}(a)--\ref{fig:eight_panel_moves}(h) plot the number of routed atom move events versus qubit count under the
partitioned architecture model and compilation strategies used in
Fig.~\ref{fig:eight_panel_fidelity}. Because native multiqubit blocks execute in
the shared entanglement zone, this metric reports the transport overhead paid
to gather gate supports across storage partitions.}
\label{fig:eight_panel_moves}
\end{figure*}

\begin{figure*}[!t]
\centering
\includegraphics[width=0.88\textwidth]{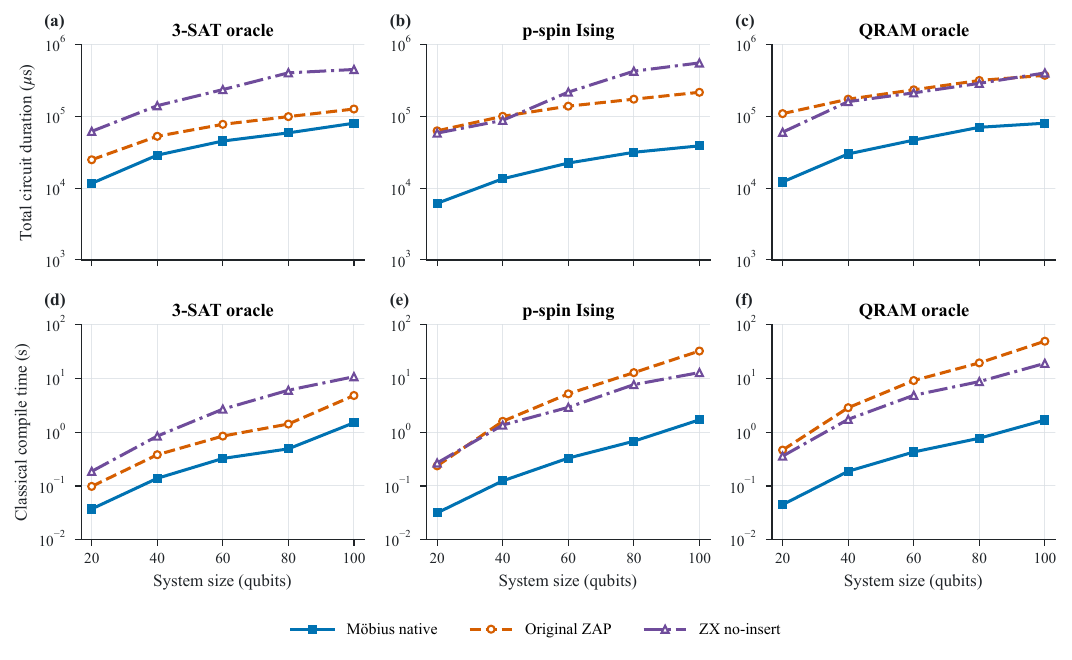}
\caption{Representative runtime-scaling comparison for three many-body/oracle
algorithm families from 20 to 100 qubits.
Figs.~\ref{fig:duration_compile_scaling}(a)--\ref{fig:duration_compile_scaling}(c)
plot total routed circuit duration versus system size.
Figs.~\ref{fig:duration_compile_scaling}(d)--\ref{fig:duration_compile_scaling}(f)
plots the corresponding total classical compile time, including decomposition
or synthesis together with neutral-atom scheduling and routing.
Figs.~\ref{fig:duration_compile_scaling}(a)--\ref{fig:duration_compile_scaling}(f)
compare M\"obius-native, ZAP-decomposed, and ZX no-insert on the same
storage-partitioned architecture.}
\label{fig:duration_compile_scaling}
\end{figure*}

All numerical comparisons use the same benchmark definitions, routing
constraints, operation durations, and no-fault fidelity factors for the three
strategies.  Figures~\ref{fig:eight_panel_fidelity} and
\ref{fig:eight_panel_moves}, together with the scheduled-stage diagnostic in Appendix~\ref{app:stage_diagnostics}, are obtained from the same routed
benchmark ensemble, while Fig.~\ref{fig:duration_compile_scaling} extends
representative many-body benchmark families to larger system sizes using the same timing
model.  In Fig.~\ref{fig:native_error_phase_diagrams}, the routed realizations are held
fixed and only the assumed native error probabilities $p_3$ and $p_4$ are
varied.

\subsection{Success rate $P_0$ across benchmark
families}

Figure~\ref{fig:eight_panel_fidelity} is the main routed \(P_0\)
comparison. The vertical axis is \(P_0\) from
Eq.~\eqref{eq:routed_no_fault}, plotted on a logarithmic scale, so a vertical
separation corresponds to a multiplicative change in the routed fidelity
estimate. Figs.~\ref{fig:eight_panel_fidelity}(a)--\ref{fig:eight_panel_fidelity}(f)
are many-body Hamiltonian or oracle cases. In these cases the M\"obius transform
exposes degree-three and degree-four occupation-projector hyperedges, and the
M\"obius-native compiler can replace long decomposed one-/two-qubit structures
by routed native multiqubit gates. This is why the M\"obius-native curve stays above the ZAP and
ZX no-insert baselines in the 3-SAT, QAOA, \(p\)-spin Ising, hypergraph, QRAM,
and multiplier cases.

The size dependence is also informative. As \(n\) grows, decomposed
baselines accumulate more two-qubit gates, routing stages, atom transfers, and
idle exposure. Their \(P_0\) values decrease rapidly. The native
path also loses \(P_0\) with system size, but it often has a gentler slope
because a single native support replaces an entire decomposed ladder. The
effect is strongest for QRAM and 4-local hypergraph instances, where the
decomposed baselines pay a large routed cost for high-control structure. It is
more moderate for the multiplier oracle because the benchmark family contains
a mixture of arithmetic components, not all of which are converted into
low-degree native gates under the \(k_{\rm nat}^{\max}=4\) table.

Figs.~\ref{fig:eight_panel_fidelity}(g) and~\ref{fig:eight_panel_fidelity}(h)
are the two-body controls. The QFT phase layer is built from pairwise controlled
phases, and the GHZ chain is essentially one- and two-qubit structure. Since no
useful degree-three or degree-four M\"obius hyperedges exist in these cases, the
M\"obius-native path has no special native resource to exploit. The
near-overlap of the three curves is therefore a control for the gains in
Figs.~\ref{fig:eight_panel_fidelity}(a)--\ref{fig:eight_panel_fidelity}(f),
where many-body diagonal terms are present.And the advantage comes from these structure.

\subsection{Atom-movement comparison}

Figure~\ref{fig:eight_panel_moves} compares the transport cost of the three
routed strategies through their atom-move counts. Native multiqubit gates
gather three or four atoms in the shared entanglement zone. For most many-body
benchmark families, however, the M\"obius-native route requires fewer or
comparable move events than the baselines because it eliminates repeated
movement associated with decomposed two-qubit ladders. The QFT and GHZ controls
overlap because all strategies route essentially the same pairwise structure.
The movement and \(P_0\) improvements occur together.
Appendix~\ref{app:stage_diagnostics} gives the companion scheduled-stage
diagnostic.

\subsection{Duration and classical compile-time scaling}

Figure~\ref{fig:duration_compile_scaling} extends the comparison to larger
systems using timing quantities rather than final \(P_0\). Figs.~\ref{fig:duration_compile_scaling}(a)--\ref{fig:duration_compile_scaling}(c) plot total routed circuit duration for 3-SAT, $p$-spin Ising, and QRAM from 20 to 100 qubits.  These benchmark families were chosen because they represent three
different many-body sources: Boolean clause oracles, local diagonal Hamiltonian terms, and multi-control addressing.  In all three cases, M\"obius-native compilation gives shorter routed duration.  The reason is the same mechanism seen in Fig.~\ref{fig:eight_panel_moves} and
Appendix~\ref{app:stage_diagnostics}: the native path reduces serialized
decomposed work and often avoids repeated movement into the
entanglement zone.  The duration advantage is particularly clear for $p$-spin Ising and QRAM, where high-degree phase
structure is frequent enough that lowering it early creates a large routed
schedule.

Figs.~\ref{fig:duration_compile_scaling}(d)--\ref{fig:duration_compile_scaling}(f)
report classical compile time, including the
decomposition/synthesis step and the neutral-atom scheduling and routing step.  This
metric is important because a useful compiler must preserve native opportunities
without replacing quantum execution cost by prohibitive classical preprocessing.
The M\"obius-native path remains faster in these experiments because it keeps a
compact hyperedge stream and avoids expanding every many-body term into a long
one-/two-qubit support list before routing.  The ZAP and ZX baselines can still
be effective circuit-generation tools, but once high-degree diagonal structure is
expanded, the routed problem contains many more elementary supports, whereas Fig.~\ref{fig:duration_compile_scaling} shows that the resource and
classical-runtime trends remain visible up to 100+ qubits.

\subsection{Strategy-level routed log-cost criterion and native-error regimes}

\begin{figure*}[!t]
\centering
\includegraphics[width=0.88\textwidth]{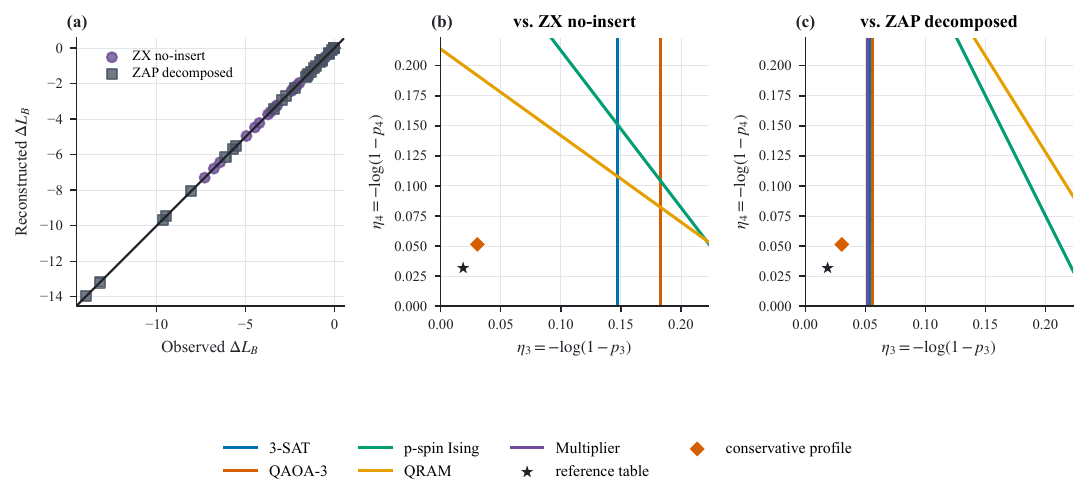}
\caption{Routed log-cost comparison.
Fig.~\ref{fig:native_error_phase_diagrams}(a) compares the count-based
reconstruction of \(\Delta L_B\) with the direct value for the 64
main-benchmark pairs; the maximum residual is \(1.8\times10^{-15}\).
Figures~\ref{fig:native_error_phase_diagrams}(b) and~\ref{fig:native_error_phase_diagrams}(c) show the \(n=30\) break-even boundaries for M\"obius-native against ZX no-insert and ZAP, respectively, as the assumed native three- and four-qubit error probabilities vary. The solid curve is \(\Delta L_B=0\); for each displayed boundary, its lower-left side has \(\Delta L_B<0\) (M\"obius-native has better performance), whereas its upper-right side has \(\Delta L_B>0\) (the corresponding baseline has better performance). The star is the reference profile and the diamond is an explicit conservative-native profile \((p_3,p_4)=(0.03,0.05)\). The 3-SAT boundary is \(p_4\)-independent because the routed native stream has only degree-three exposure. In Fig.~\ref{fig:native_error_phase_diagrams}(b), Multiplier has \(\Delta N_3=\Delta N_4=0\) against ZX no-insert, so its \(\Delta L_B=-0.253\) is independent of \((p_3,p_4)\) and no break-even curve appears.}
\label{fig:native_error_phase_diagrams}
\end{figure*}

Figure~\ref{fig:native_error_phase_diagrams} verifies the routed
log-cost reconstruction in Fig.~\ref{fig:native_error_phase_diagrams}(a) and then shows how the comparison changes with the
assumed three- and four-qubit native error probabilities.

\begin{samepage}
Let \(L_s=-\log P_{0,s}\) for routed strategy \(s\), and let \(B\) denote
a baseline. The routed gate and event counts give the exact relation
\begin{equation}
\Delta L_B \equiv L_{\rm M}-L_B=\Delta A_B+\Delta N_{3,B}\eta_3+\Delta N_{4,B}\eta_4,
\end{equation}
\end{samepage}

where \(\eta_k = -\log F_{\rm nat}^{(k)}\). Here \(\Delta A_B\) contains every shared-profile contribution other than
the native three- and four-qubit factors, and \(\Delta N_{k,B}\) is the routed
count difference. Thus M\"obius-native has larger \(P_0\) exactly when
\(\Delta L_B<0\).

Figs.~\ref{fig:native_error_phase_diagrams}(b) and
\ref{fig:native_error_phase_diagrams}(c) sweep
\(p_k=1-F_{\rm nat}^{(k)}\) for the routed \(n=30\) streams. The 3-SAT oracle contains only degree-three native candidates under the benchmark construction, so its break-even boundary is almost independent of $p_4$ and appears as a vertical line.  This is the expected behavior for a case whose native advantage comes only from $CCZ$-type blocks.  The $p$-spin Ising and QRAM instances contain both degree-three and degree-four candidates, so their boundaries bend across the $(p_3,p_4)$ plane.  In these panels a poor four-qubit native gate can remove the M\"obius advantage even if three-qubit gates are good, and vice versa. The comparison also depends on the baseline: ZX no-insert and ZAP have different decomposed gate streams, routed stages, and movement patterns, so the break-even contours are not identical.

\section{Discussion}
\label{sec:discussion}

The main implication of this work is that multiqubit diagonal structure should be treated as a compilation resource rather than an intermediate form to be eliminated at the beginning of compilation. In the benchmark families studied here, the M\"obius spectrum identifies when a target diagonal unitary contains genuine high-degree occupation-projector phases, while the routed neutral-atom evaluation determines whether those phases remain useful after movement, scheduling, and native-gate errors are included. This separation is essential: a high-degree phase term is an opportunity for native execution, but its value depends on whether the participating atoms can be gathered and executed with less total cost than a decomposed alternative.

This view also places M\"obius-native compilation in relation to decomposed
and ZX-based streams. Its purpose is not to prescribe native multiqubit
execution in every instance, but to preserve many-body phase structure until
the target native gate table is applied. When the diagonal structure is only
low-degree, when the relevant hyperedges are geometrically unfavorable, or when
a decomposed or ZX-derived phase circuit is more compact under the same hardware
assumptions, lowering to one- and two-qubit gates may be preferable. The
contribution of the M\"obius representation is therefore to make a routed,
hardware-aware comparison possible instead of eliminating the candidate
supports at the front end.

The numerical model used here should be viewed as a controlled
compiler-level testbed for comparing routed schedules under different
native-gate assumptions. Our purpose is to expose when preserving high-degree diagonal structure can improve movement, timing, and \(P_0\). Because the
M\"obius phase-hypergraph representation is independent of the particular proxy
error table, the same framework can incorporate device-specific native
fidelities, geometry predicates, movement times, atom-loss data, crosstalk maps, and stronger routing objectives in future neutral-atom studies.

The same compilation question is also relevant beyond the physical-level diagonal benchmarks considered here. Fault-tolerant and error-diagnostic protocols often contain structured many-qubit phases, parity conditions, or feedback-controlled operations that can either be executed as native coherent events or lowered into conventional gate sequences. Stabilizer extraction, syndrome-dependent feedback, measurement-free correction, lattice-surgery-type primitives, and dynamically generated logical memories all include operations that couple a small set of qubits through a parity, projector, or controlled-phase condition~\cite{Dennis2002,Fowler2012,Cong2022,Perlin2023,Hastings2021,Gidney2021,Bluvstein2024}. Recent work on multiqubit Rydberg gates for QEC makes this connection explicit: Locher \emph{et al.} show how CCZ-type gates and global three-qubit Rydberg gates can support measurement-free QEC and Floquet-code stabilizer readout while reducing circuit depth and shuttling overhead in neutral-atom implementations~\cite{Locher2026}. In such settings, the M\"obius phase-hypergraph representation could provide a natural interface for exposing diagonal stabilizer phases, syndrome projectors, flag phases, and controlled-feedback phases before the compiler commits to either native execution or decomposition.

Several extensions follow naturally from this framework. A device-specific implementation could replace the representative native-error parameters by support-dependent fidelities measured for different atom configurations, blockade radii, Rydberg pulse families, and local geometries. The routing layer could be upgraded from a greedy scheduler to a global or hybrid optimizer that trades off native fan-in, atom motion, parallelism, idle exposure, and crosstalk. At the logical level, the same IR could be coupled to decoder-aware objectives, correlated-error models, QEC-cycle-level cost functions, and logical reliability metrics. In all of these directions, the central design principle remains the same: many-body diagonal structure should remain explicit long enough for architecture- and calibration-dependent compilation decisions to be made.

\section{Conclusion}
\label{sec:conclusion}

We have presented a structure-preserving compilation framework for
diagonal neutral-atom algorithms based on a M\"obius phase-hypergraph
intermediate representation. M\"obius inversion converts basis-state phases,
phase oracles, and local diagonal Hamiltonians into occupation-projector
hyperedges, allowing many-body diagonal structure to remain explicit throughout
hardware-aware compilation. This representation is especially natural for
neutral-atom processors, where Rydberg-mediated interactions, atom motion,
storage partitions, and entangling-zone constraints make the relation between
algebraic circuit structure and physical scheduling a central compilation
problem.
Taken together, the results show that many-body diagonal phases should be
treated as a schedulable architecture-level resource rather than as structure
to be immediately decomposed. The role of the M\"obius phase hypergraph is to
keep this resource visible until routing cost, zone capacity, idle exposure,
crosstalk, timing, and calibrated native-gate errors can be assessed jointly.
This shifts the compiler objective from minimizing an abstract gate count to
making calibration-driven architectural choices. It also provides a natural
path toward future neutral-atom compilers for QEC and logical algorithms, where
multiqubit check operations, atom-loss handling, decoder-aware scheduling, and
logical-level cost functions must be optimized within a common structure-preserving framework.

\begin{acknowledgments}
The authors thank Chen Qian, Weifeng Zhuang, Hongze Xu, and Xudan Chai for helpful discussions. This work was supported by the National Key Research and Development Program of China (Grant No. 2025YFE0217200), the National Natural Science Foundation of China (Grant No. 92365111 and No. 12404559), and Shanghai Municipal Science and Technology (Grant No. 25LZ2600200).
\end{acknowledgments}

\appendix

\section{Möbius inversion on the subset lattice}
\label{app:mobius-inversion}

In this appendix we briefly review the Möbius inversion formula used in
the main text. Let \([n]=\{1,\ldots,n\}\), and let \(2^{[n]}\) denote the
Boolean lattice of all subsets of \([n]\), ordered by set inclusion. For
two functions
\[
    f,g:2^{[n]}\rightarrow \mathbb{R},
\]
the subset-zeta transform is the map
\[
    g(T)=\sum_{S\subseteq T} f(S),
    \qquad T\subseteq [n].
\]
In words, \(g(T)\) is obtained by summing the values of \(f(S)\) over all
subsets \(S\) contained in \(T\). Möbius inversion states that this map is
invertible, and its inverse is
\[
    f(S)=\sum_{R\subseteq S}(-1)^{|S|-|R|}g(R),
    \qquad S\subseteq [n].
\]
This is the finite-set version of the inclusion-exclusion principle.

In our setting, the role of \(g(T)\) is played by the computational-basis
phase \(F(T)\), and the role of \(f(S)\) is played by the native projector
phase \(\theta_S\). The native phase-accumulation rule gives
\[
    F(T)=\sum_{S\subseteq T}\theta_S.
\]
Therefore, by Möbius inversion,
\[
    \theta_S
    =
    \sum_{R\subseteq S}
    (-1)^{|S|-|R|}F(R).
\]
This is Eq.~\eqref{eq:thetas} in the main text.

Index the vectors \(F\) and \(\theta\) by subsets and define
\(\zeta_{T,S}=\mathbf 1[S\subseteq T]\). Then \(F=\zeta\theta\), and the
inverse has entries
\[
M_{S,R}=(-1)^{|S|-|R|}\mathbf 1[R\subseteq S],
\qquad M=\zeta^{-1}.
\]
In product bit-string ordering, these transforms factor as
\begin{equation}
\begin{aligned}
\zeta&=\zeta_1^{\otimes n},&
M&=\mu_1^{\otimes n},\\
\zeta_1&=
\begin{pmatrix}
1&0\\
1&1
\end{pmatrix}, &
\mu_1&=
\begin{pmatrix}
1&0\\
-1&1
\end{pmatrix}.
\end{aligned}
\label{eq:appendix_kronecker}
\end{equation}
Since \(\mu_1\zeta_1=I_2\), tensoring gives \(M\zeta=I_{2^n}\). Let
\(\Pi\) reorder subsets by degree, as in
Fig.~\ref{fig:native_mobius_overview}(c). The displayed matrix is
\(\Pi M\Pi^{-1}\). Its entries vanish unless the column support is contained
in the row support, which gives the block lower-triangular form and the
diagonal block \(I_{(k)}\) of size \(\binom{n}{k}\).

We now give the proof. Starting from the proposed inverse formula and
substituting \(F(R)=\sum_{U\subseteq R}\theta_U\), we obtain
\[
\begin{aligned}
    \sum_{R\subseteq S}(-1)^{|S|-|R|}F(R)
    &=
    \sum_{R\subseteq S}(-1)^{|S|-|R|}
    \sum_{U\subseteq R}\theta_U  \\
    &=
    \sum_{U\subseteq S}\theta_U
    \sum_{R:\,U\subseteq R\subseteq S}
    (-1)^{|S|-|R|}.
\end{aligned}
\]
For a fixed \(U\subseteq S\), write \(R=U\cup A\), where
\(A\subseteq S\setminus U\). Then
\[
\begin{aligned}
\sum_{R:\,U\subseteq R\subseteq S}(-1)^{|S|-|R|}
&=\sum_{A\subseteq S\setminus U}(-1)^{|S|-|U|-|A|}\\
&=(1-1)^{|S|-|U|}.
\end{aligned}
\]
This expression equals \(1\) when \(U=S\), and equals \(0\) otherwise.
Hence all terms cancel except the term \(U=S\), giving
\[
    \sum_{R\subseteq S}(-1)^{|S|-|R|}F(R)=\theta_S.
\]
Thus the inverse formula exactly recovers the native projector phases
\(\theta_S\) from the computational-basis phases \(F(T)\).

For two and three qubits, the inclusion--exclusion rule gives
\begin{equation}
\begin{aligned}
\theta_{\{1,2\}}
&=F_{11}-F_{10}-F_{01}+F_{00},\\
\theta_{\{1,2,3\}}
&=F_{111}-F_{110}-F_{101}-F_{011}\\
&\quad+F_{100}+F_{010}+F_{001}-F_{000}.
\end{aligned}
\label{eq:low_degree_mobius_examples}
\end{equation}
Thus \(\theta_{\{1,2\}}\) is the two-body phase
contribution after the global
and one-body terms have been removed, and \(\theta_{\{1,2,3\}}\) is the
three-body contribution after all proper-subset terms have been removed.

Since physical diagonal gates are defined only up to phases modulo
\(2\pi\), the equalities above should be understood for any chosen real
lift of the phases \(F(T)\). Different choices of representatives may
change some \(\theta_S\) by integer multiples of \(2\pi\), but they produce
the same diagonal unitary. If global phases are ignored, one may set
\(F(\varnothing)=0\) and omit the term \(\theta_\varnothing\).

\section{Locality proof for the output-sensitive M\"obius bound}
\label{app:local_complexity_proof}

Building on the subset-lattice inversion reviewed in
Appendix~\ref{app:mobius-inversion}, we justify
Eq.~\eqref{eq:local_complexity}.  Write the M\"obius coefficient of a
diagonal phase function $F$ as
\begin{equation}
\theta_S[F]=\sum_{R\subseteq S}(-1)^{|S|-|R|}F(1_R),
\label{eq:appendix_mobius_coeff}
\end{equation}
where $1_R$ is the bit string whose occupied set is $R$. The map
$F\mapsto\theta_S[F]$ is linear. For the decomposition in Eq.~\eqref{eq:local_phase_decomposition},
\begin{equation}
\theta_S[F]=\sum_{\alpha=1}^{N_{\rm loc}}
\theta_S[f_\alpha].
\label{eq:appendix_linearity}
\end{equation}

A local term \(f_\alpha(x_{A_\alpha})\) has no M\"obius coefficient on a
support outside \(A_\alpha\). If
\(j\in S\setminus A_\alpha\), the two contributions in
Eq.~\eqref{eq:appendix_mobius_coeff} from each pair $R$ and $R\cup\{j\}$ cancel,
because \(f_\alpha\) is independent of \(x_j\). Hence
\(\theta_S[f_\alpha]=0\) unless
\(S\subseteq A_\alpha\).

Each transform can therefore be restricted to the \(|A_\alpha|\le k\)
variables of one local term. A fast subset M\"obius transform on that table
costs \(O(k2^k)\) time and \(O(2^k)\) memory and yields at most \(2^k-1\)
nonconstant supports. Summing over \(N_{\rm loc}\) terms gives
\(O(N_{\rm loc}k2^k)\) arithmetic work and at most
\(O(N_{\rm loc}2^k)\) generated
supports before duplicate supports are merged or coefficients vanish modulo
$2\pi$.  This proves the output-sensitive local bound used in the main text.

For an explicit example, encode the field register at site \(i\) as
\(\phi_i=\phi_0+\sum_{b=0}^{m-1}w_bx_{i,b}\), with \(N\) sites and \(m\)
bits per site, so \(n_{\rm field}\approx Nm\). Using
\(x_{i,b}^2=x_{i,b}\), an onsite polynomial \(V(\phi_i)\) of fixed degree
\(d\) contains at most \(O(m^d)\) distinct bit monomials per site and hence
produces \(O(Nm^d)\) projector phases. A nearest-neighbor quadratic term
\(\sum_{\langle i,j\rangle}(\phi_i-\phi_j)^2\) produces
\(O(N_{\rm bond}m^2)\) phases over \(N_{\rm bond}\) lattice bonds. Both
contributions are polynomial in \(N\) and \(m\) for fixed \(d\), as required
by Eq.~\eqref{eq:local_complexity}.

\clearpage
\onecolumngrid
\clearpage

\section{Additional benchmark landscape}
\label{app:additional_landscape}

Table~\ref{tab:additional_landscape} lists additional families
from an auxiliary $n=100$ no-insert comparison, using the same ``input supports /
routed stages'' reporting format to separate the M\"obius-native, ZAP-decomposed,
and ZX no-insert outcomes.

\begin{widetext}
\refstepcounter{table}
\label{tab:additional_landscape}
\footnotesize
\textbf{TABLE~\thetable.} Additional benchmark families not plotted in the main
comparison figures.  ``Deg.'' is the largest relevant M\"obius projector support
in the benchmark construction.  The three method columns report input support
count / routed stage count at $n=100$; smaller values are better.
\vspace{0.5em}
\begin{center}
\begin{ruledtabular}
\begin{tabular}{p{0.16\textwidth}p{0.23\textwidth}c c c c p{0.17\textwidth}}
Family & Representative structure & Deg. &
M\"obius\HHR{-}native & ZAP\HHR{-}decomposed & ZX no-insert & Effect \\
\hline
Grover-4 oracle & Patterned 4-local Grover oracle blocks & 4 & $484/133$ & $7564/3043$ & timeout ($>1$h) & Native stages are $22.9\times$ fewer than ZAP \\
QPE phase layer & Pairwise controlled-phase estimation layer & 2 & $660/68$ & $660/68$ & $660/287$ & Two-body control; no native advantage over ZAP \\
Bernstein--Vazirani/DJ & Parity-oracle phase layer & 2 & $99/99$ & $99/99$ & $99/99$ & All three paths coincide \\
MaxCut QAOA & 2-local Ising cost layer & 2 & $148/26$ & $148/26$ & $148/29$ & Mostly a two-body control \\
VQE-HWE / VQC & Hardware-efficient trainable ansatz & 2 & $150/148$ & $150/148$ & $150/149$ & Base ansatz has no high-degree native target \\
\end{tabular}
\end{ruledtabular}
\end{center}
\end{widetext}

\clearpage
\onecolumngrid
\clearpage

\section{Scheduled-stage diagnostics}
\label{app:stage_diagnostics}

The main text reports \(P_0\) and
atom-movement cost as the primary
end-to-end quantities.  The scheduled-stage count is kept here as a companion
diagnostic because it isolates how much serialized routed work remains after
scheduling.

\begin{figure}[!ht]
\centering
\includegraphics[width=0.94\textwidth]{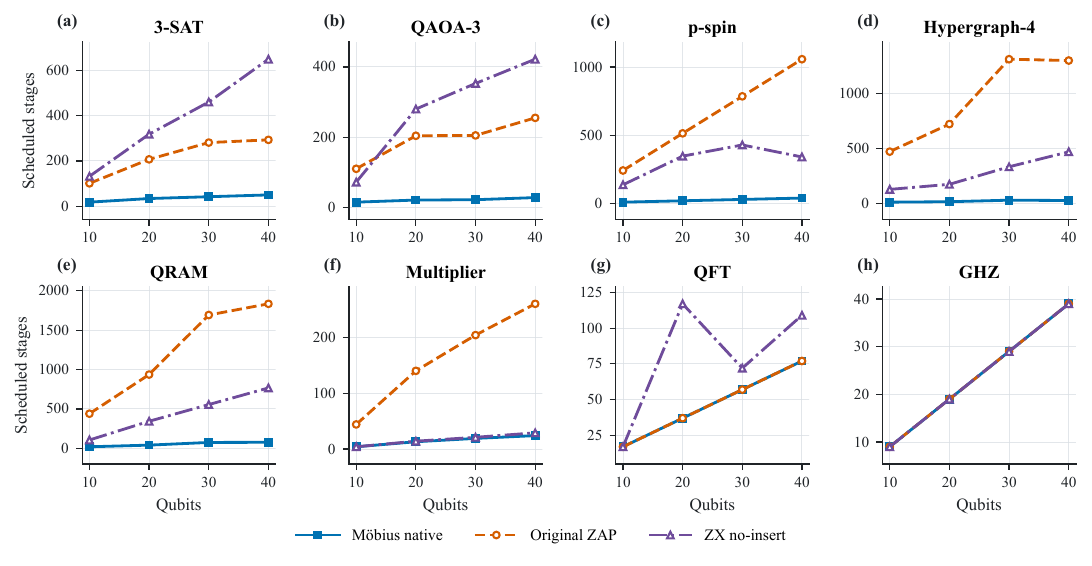}
\caption{Scheduled-stage comparison for eight common algorithmic benchmark
families. Figs.~\ref{fig:eight_panel_stages}(a)--\ref{fig:eight_panel_stages}(h)
plot the routed ASAP scheduled stage count versus qubit count for the benchmark
families and compilation strategies used in
Figs.~\ref{fig:eight_panel_fidelity} and~\ref{fig:eight_panel_moves}.
Figs.~\ref{fig:eight_panel_stages}(a)--\ref{fig:eight_panel_stages}(f) contain
explicit many-body Hamiltonian or oracle terms and expose the stage reduction
available from native multiqubit execution.
Figs.~\ref{fig:eight_panel_stages}(g) and~\ref{fig:eight_panel_stages}(h) are
two-body controls where the M\"obius-native and ZAP-decomposed paths largely
coincide.}
\label{fig:eight_panel_stages}
\end{figure}

Figure~\ref{fig:eight_panel_stages} reports the scheduled-stage count after the
common routed ASAP scheduler. A stage is a
layer of mutually compatible routed operations subject to qubit conflicts, shared
entanglement-zone capacity, and movement-layer constraints.

\clearpage

\section{ZX insert comparison}
\label{app:zx_insert_diagnostic}

In the main text, Figure~\ref{fig:eight_panel_fidelity} and~\ref{fig:eight_panel_moves} employ the ZX no-insert mode, our consistent conservative ZX baseline for all routed benchmark evaluations. Under this paradigm, source circuits are simplified and extracted without introducing extra ZX graph structures. The ZX-insert mode, by contrast, implements a more aggressive optimization scheme. It enables the ZX extraction pipeline to embed supplementary graph components prior to extraction, which can reveal distinct multi-controlled phase blocks and reduce the plotted gate count for certain benchmark instances. The ZX-insert mode serves as a practical diagnostic tool for evaluating performance differences across compilation algorithms.

\begin{figure}[!ht]
\centering
\includegraphics[width=0.88\textwidth]{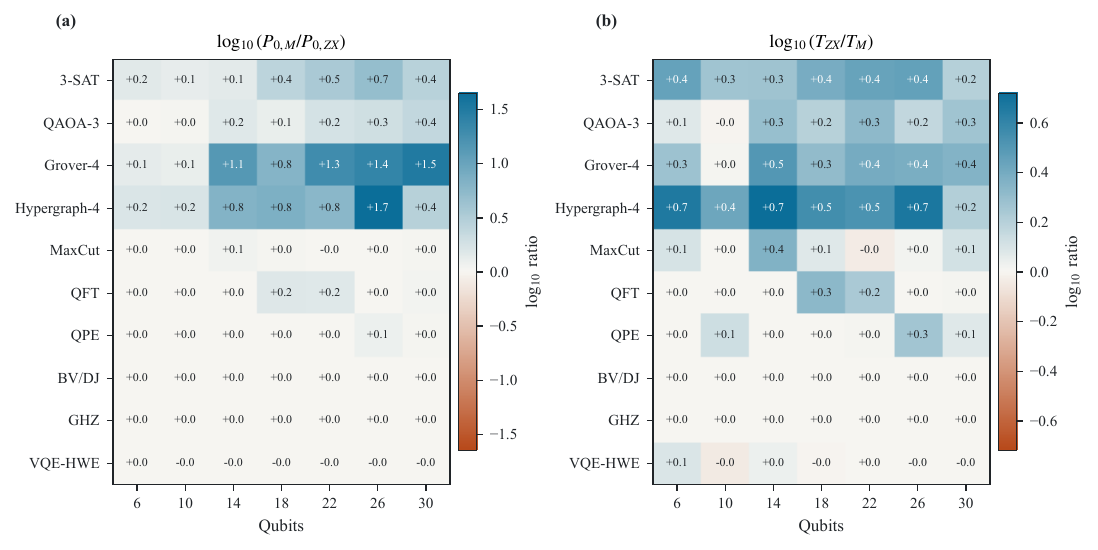}
\caption{Routed ZX-insert comparison, with seven
qubit counts in each of ten benchmark families.
Fig.~\ref{fig:zx_insert_full_audit}(a) reports
\(\log_{10}(P_{0,\mathrm{M}}/P_{0,\mathrm{ZX}})\), where positive values favor
M\"obius-native in routed \(P_0\).
Fig.~\ref{fig:zx_insert_full_audit}(b) reports
\(\log_{10}(T_{\mathrm{ZX}}/T_{\mathrm{M}})\), where positive values indicate
shorter M\"obius-native routed duration.}
\label{fig:zx_insert_full_audit}
\end{figure}

\begin{figure}[!ht]
\centering
\includegraphics[width=0.94\textwidth]{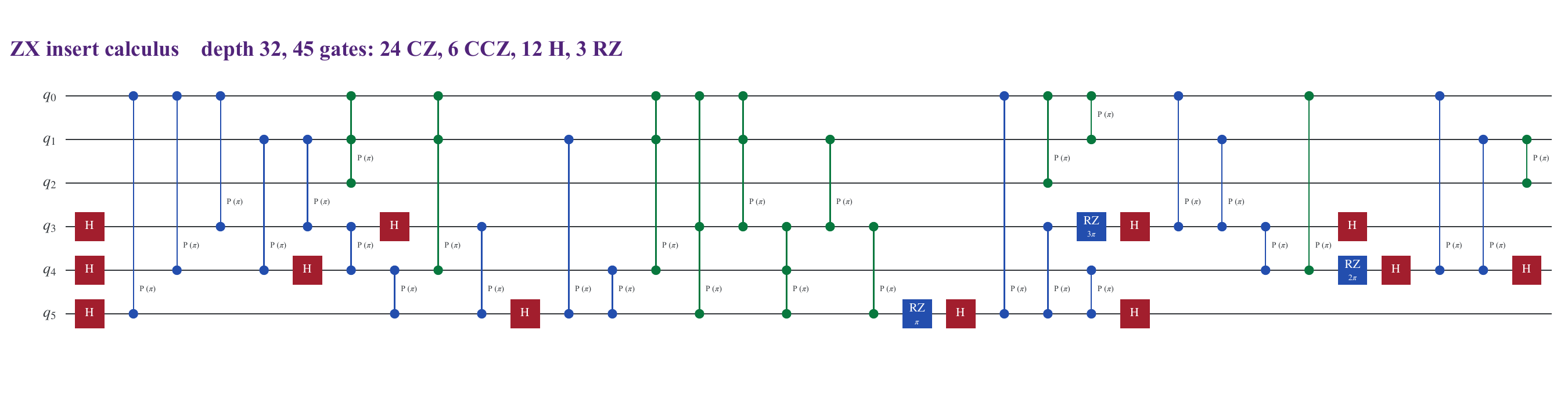}
\caption{Auxiliary ZX insert circuit for the same compact 6-qubit 3-SAT oracle instance used in Fig.~\ref{fig:sat3_circuit_comparison}. The circuit is drawn with the same gate notation as Fig.~\ref{fig:sat3_circuit_comparison}: filled dots mark the participating qubit lines of controlled-phase gates; each multiqubit projector phase is labeled \(P(\theta)\) on its exact filled-dot support, and RZ angles are shown in parentheses next to the RZ gate. For this instance, the displayed ZX insert circuit has depth \(32\) and \(45\) displayed gates, including six displayed \(CCZ\) gates, whereas Fig.~\ref{fig:sat3_circuit_comparison}(b) has displayed depth \(37\) and \(64\) displayed gates with one displayed \(CCZ\) gate.}
\label{fig:sat3_circuit_comparison_zx_insert}
\end{figure}

Fig.~\ref{fig:zx_insert_full_audit} reports the routed comparison over 70
source instances, and Fig.~\ref{fig:sat3_circuit_comparison_zx_insert} gives the
insert-mode circuit for the representative six-qubit 3-SAT instance.

The larger positive ratios for Grover-4 and Hypergraph-4, together with the near-zero ratios for the two-body controls, are consistent with the main finding that the routed value of phase-structure preservation depends on the benchmark representation.

Figure~\ref{fig:sat3_circuit_comparison_zx_insert} shows the ZX insert
circuit for the same compact six-qubit 3-SAT instance. The insert mode
shortens the ZX circuit in this instance, reducing the displayed ZX depth/gate
count from $37/64$ in Fig.~\ref{fig:sat3_circuit_comparison}(b) to $32/45$ here
and exposing six displayed $CCZ$ blocks.  The resulting circuit should therefore
be interpreted as a stronger ZX front-end variant, not as the fixed ZX no-insert
reference used in the main routed comparisons.

\clearpage
\twocolumngrid

\clearpage

\bibliography{references}

\end{document}